# Deep-sub-cycle ultrafast optical pulses

**Hongliang Dang,[+] Jiaxin Gao,[+] Hao Wu, Xin Guo & Limin Tong[*]**

*New Cornerstone Science Laboratory, State Key Laboratory of Extreme Photonics and Instrumentation, College of Optical Science and Engineering, Zhejiang University, Hangzhou 310027, China*

[+] These authors contribute equally.

* Corresponding author: phytong@zju.edu.cn

## Abstract

**Sub-cycle optical pulse is of great importance for ultrafast science and technology. While a narrower pulse can offer a higher temporal resolution, so far the pulse width has not reached the limit of half an optical cycle. Here we propose to break the half-cycle limit via inverse Compton scattering in a deep-subwavelength-confined infrared or THz optical driving field, which can be converted into a deep-sub-cycle (i.e., less than half a cycle) ultrafast pulse by relativistic electrons passing through. Quantitatively, by using a deep-subwavelength-confined 0.4-THz 1-ps pulsed driving field and a 3-MeV 1-fs electron bunch, we obtain a ~0.1-cycle femtosecond pulse with a peak frequency of ~26 THz (~38 fs in optical cycle) and a pulse width of ~3.6 fs; with the same driving field and a 30-MeV 50-as electron bunch, we obtain a 0.17-cycle attosecond pulse with a peak frequency of ~2.0 PHz (~500 as in optical cycle) and a pulse width of ~86 as. Below the optical damage threshold of the field-confinement material, the single-pulse photon number can exceed $10^6$. Such pulses may open opportunities for studying light-matter interaction on the deep-sub-cycle level, and pave a way to unprecedented optical technology ranging from temporal super-resolution optical microscopy and spectroscopy to unconventional atom/molecule polarization and manipulation.**

Space and time are basic scales for understanding the world. Higher spatial (e.g., from optical[1,2] to electron microscopy[3,4]) and temporal (e.g., from femtosecond[5,6] to attosecond[7,8] technology) resolutions are always desired for a better understanding of the microscopic world on a deeper level[5,8]. On the temporal scale, so far the highest resolution comes from ultrafast optical technology, which has been widely employed for studying the dynamics of molecular or atomic processes from femtosecond to attosecond scales[3,5-10]. Typically, to pursue a higher temporal resolution, we have to resort to shorter optical pulses with higher photon energy (i.e., shorter wavelength), which is analogous to achieving higher diffraction-limited spatial resolution by reducing the wavelength of the probing light (e.g., X-ray microscopy[11]). However, for a certain light-matter interaction, reducing the wavelength far away from the intrinsic resonance of the matter will significantly reduce the excitation efficiency (i.e., the matter becomes “transparent” to the probing light). Therefore, a more effective way to increase the spatial resolution and resolve the sub-diffraction structure is not solely by reducing the wavelength, but by confining the light field to a spatial scale well below the half wavelength (i.e., deep-subwavelength scale) that breaks the optical diffraction limit in the near field, and obtain a super-diffraction-limit optical resolution at a wavelength not much far away from the concerned resonance of the material, such as optical near-field microscopy [1,12,13]. Compared with the spatial super-resolution technology that has been well developed, the temporal super-resolution remains challenging, due to the absence of the deep-sub-cycle optical field with a pulse width shorter than half a cycle.

In recent years, various approaches[14-21] based on e.g., optical parametric chirped pulse amplification (OPCPA)[14,15], optical arbitrary waveform generation (OAWG)[16,17], and relativistic electron sheet (RES)[18,19], have been developed to generate sub-cycle ultrafast optical pulses. However, due to difficulties such as limited gain bandwidth of femtosecond laser sources[20] and insufficient degree of freedom of waveform control[21], the narrowest pulse width obtained remains larger than $T_c/2$ ($T_c/2$ on the temporal scale is analogous to the diffraction limit of $\lambda/2$ on the spatial scale, where $T_c$ and $\lambda$ are the cycle and the wavelength of the light).

Here, we propose to generate deep-sub-cycle ultrafast optical pulses via inverse Compton scattering (ICS) in a deep-sub-wavelength-confined optical field, in which the field with a spatial full width at half maximum ($FWHM_S$) smaller than half a wavelength is converted into a deep-sub-cycle optical pulse with a temporal full width at half maximum ($FWHM_T$) smaller than half a cycle by relativistic electrons passing through.

**Breaking the half-cycle temporal limit**

Our approach to breaking the half-cycle limit of an optical pulse is illustrated in Fig. 1. In the laboratory coordinate system, when a photon (with a frequency $\nu_0$ and wavelength $\lambda_0$) collides with an electron with much higher energy (e.g., a free-space relativistic electron), it can be converted into a photon with higher energy (i.e., a higher frequency $\nu_1$ and a shorter wavelength $\lambda_1$) by ICS (Fig. 1a; see also section 1 in Supplemental Material). Usually, the photon is provided by a focused laser beam in free space with a spot size larger than the diffraction limit (Fig. 1b), as has been reported in previous works[22-24]. When an electron with a velocity $V_e$ passes through the focused optical field with a Gaussian-profile intensity distribution ($FWHM_S \geqslant \lambda_0/2$), the scattered photons will compose a virtual pulse (compared to a typical pulse that usually contains a large number of photons, the single-electron-scattered pulse typically has an average photon number much less than 1, and is thus called a "virtual" pulse here) with a width $\tau_1$ that can be estimated by

$$\tau_1 \approx \left(\frac{\mathrm{FWHM_S}}{V_\mathrm{e}}\right)\cdot(1-\beta\cos\theta) \tag{1}$$

where $\theta$ is the emission angle (see Fig. 1a), $\beta=V_e/c$, and the factor ($1-\beta\cos\theta$) comes from the time compression effect for the observer facing the scattered photon [25]. Since $\lambda_1=\lambda_0\cdot(1-\beta\cos\theta)$ (see section 1.1 in Supplemental Material), $V_e<c$, we have $\tau_1>\lambda_1/2c=T_{c1}/2$, showing that the width of a pulse composed of conventional ICS photons can hardly be shorter than the half-cycle limit $T_{c1}/2$.

Recently, a deep-sub-wavelength-confined optical field has been realized in a slit waveguide mode supported by a coupled nanowire pair (CNP) [26-28], which can offer a high-intensity peak field with $FWHM_S << \lambda_0/2$ at the center of the mode in the optical near field (Fig. 1c; see also section 1.2.1 in Supplemental Material), and a low-intensity diffraction-limited background field with a peak-to-background ratio up to 30 dB. When an electron passes through such a field, a deep-sub-cycle virtual pulse can be obtained with a temporal width $\tau_1<T_{c1}/2$ providing that $V_e>2c\cdot FWHM_S/\lambda_0$. For example, if $FWHM_S/\lambda_0=1/4$, to have $\tau_1<T_{c1}/2$, the electron should have $V_e > 0.5c$, which can be readily achieved by relatively small electron acceleration systems (e.g., an electron microscope [29] ).

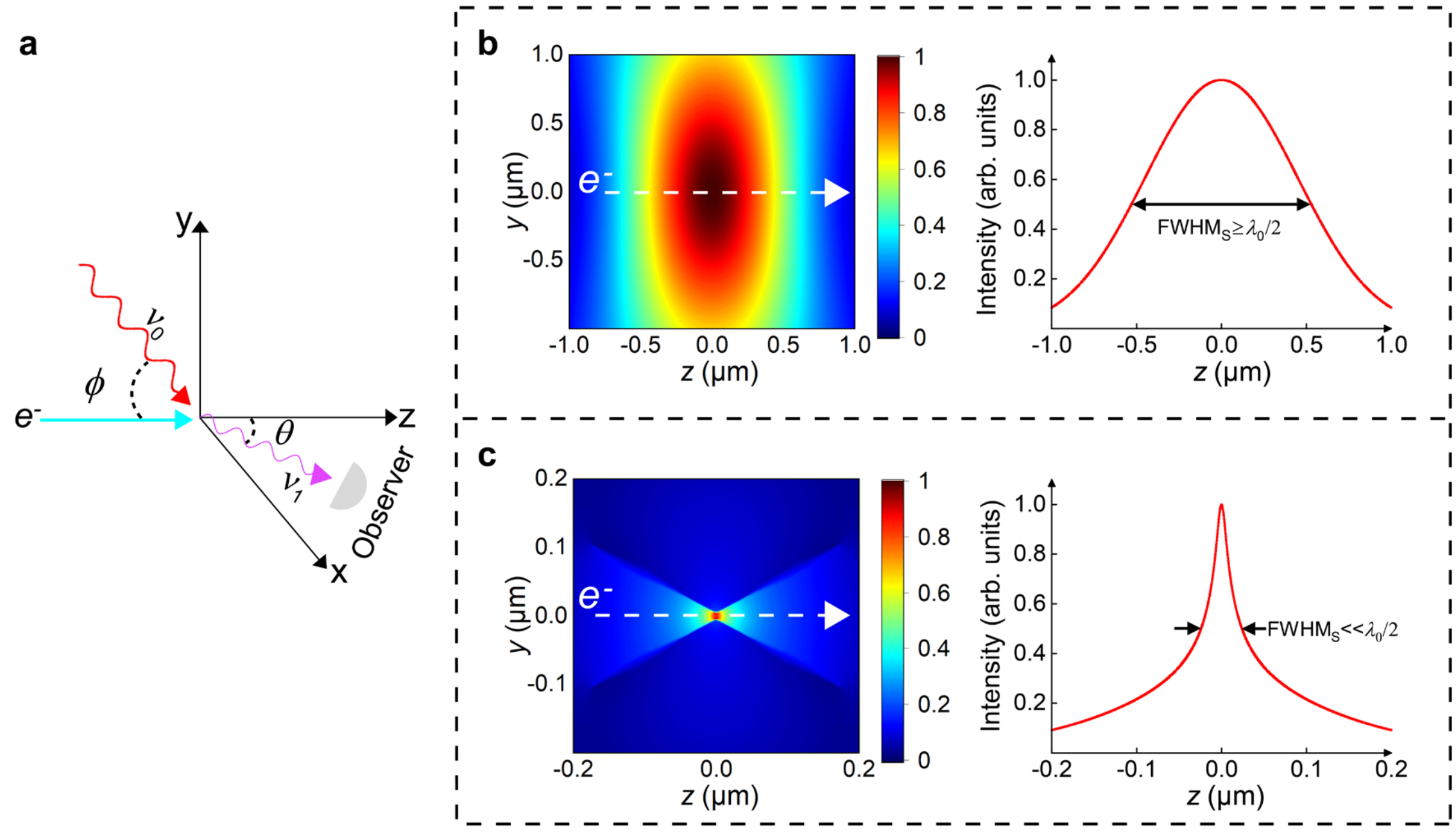

**Fig. 1| Schematic illustration of breaking the half-cycle temporal limit of an optical pulse. a,** Inverse Compton scattering (ICS) between a photon and a relativistic electron. Typically, the electron energy is much higher than the initial photon energy, and the photon energy will be increased after ICS (i.e., $\nu_1>\nu_0$), while the energy loss of the electron is negligible. **b,** Calculated field intensity distribution (left panel) and $FWHM_S$ (right panel) of a conventional Gaussian-profile 1.8-μm-wavelength optical field that is focused on approaching the diffraction limit. Due to the diffraction limit, the $FWHM_S$ cannot be smaller than $\lambda_0/2$, which sets the lower limit of the transition time ($\tau_1$) to be larger than half an optical cycle when an electron passes through. **c,** Calculated field intensity distribution (left panel) and $FWHM_S$ (right panel) of a deep-sub-wavelength confined 1.8-μm-wavelength optical field obtained in a CNP composed of two 500-nm-diameter Cadmium Sulfide (CdS) nanowires, with a slit width of 20 nm. Here the $FWHM_S$ can be much smaller than $\lambda_0/2$, making it possible to decrease $\tau_1$ well below half an optical cycle, and generate a deep-sub-cycle optical pulse that breaks the half-cycle temporal limit.

## Deep-sub-cycle virtual pulses obtained with single-electron incidence

For principle exploration, we first consider the simplest case with single-electron incidence. We assume that the wavelength of the pulsed optical driving field ($\lambda_0$) is 1.8 μm (a common wavelength of the driving field in high harmonic generation (HHG)[30,31]), and the field is confined as a slit waveguide mode formed by coupling two identical Cadmium Sulfide (CdS) single-crystal nanowires (i.e., a coupled nanowire pair, CNP, see Fig. 2a)[26,27]. By assuming a slit width of 20 nm and a nanowire diameter of 500 nm, we have an $FWHM_S$ ~77 nm (corresponding to an $FWHM_S/\lambda_0 \sim 0.04$) of the central peak along the *z*-direction (Fig. 2b). Meanwhile, we assume that the electron has a kinetic energy of 1 MeV (corresponding to a velocity $V_e$~0.94$c$), and passes through the confined optical field from left to right along the central symmetry line (i.e., the *z*-direction) of the CNP cross-section (Fig. 2a). Due to the symmetry of the CNP structure and thus that of the optical driving field, the electron collides with photons almost perpendicularly on their motion trajectories (i.e., $\phi=\pi/2$ in Fig. 1a). In such a case, the path-dependent (i.e., *z*-dependent) scattered photon number $n(z)$ can be obtained as [22] (see also section 1.2.2 in Supplemental Material for details)

$$n(z) = N_0 N_e \sigma_T \frac{S(z)}{P_{mode}} \frac{\delta_t}{\tau_0} \tag{2}$$

where the total photon number of the driving field $N_0 \approx E_{p0}/h\nu_0$, in which $E_{p0}$ is the total energy of the driving pulse, $h$ is the Planck constant, and $\nu_0$ the central frequency of the driving pulse; $N_e$ is the number of incident electrons (here $N_e$=1), $\sigma_T$ is the Thomson scattering cross section ($\sigma_T$ =6.65×10$^{-29}$ m$^2$), $S(z)$ is the $z$-dependent Poynting vector, $P_{mode}$ is the mode power (i.e., the mode area integral of Poynting vector) of the optical driving field at the moment the electron passing through, $\delta_t$ is the transition time of the electron, and $\tau_0$ is the width of the driving pulse (with an assumption of $\tau_0 > \delta_t$). Since $\delta_t$ (typically tens to hundreds of attoseconds) is much shorter than $\tau_0$ (typically tens to hundreds of femtoseconds), $P_{mode}$ is almost a constant during the time of electron transit, and can be assumed as the average power of the pulsed driving field (i.e., $P_{mode}=E_{p0}/\tau_0$).

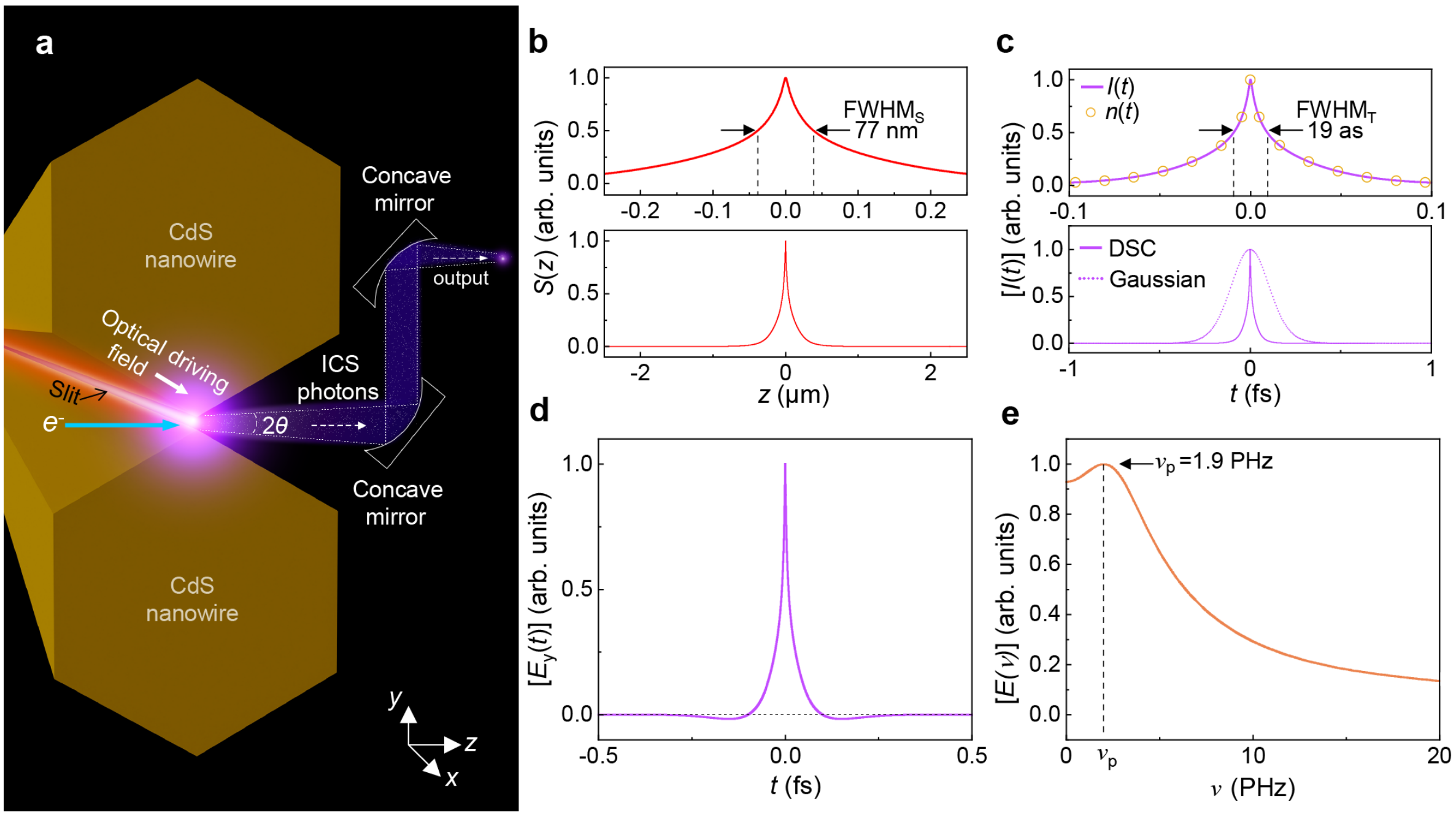

**Fig. 2| Generation of deep-sub-cycle optical pulses with single-electron incidence. a,** Schematic diagram of the ICS system. An electron incident from the left side collides perpendicularly with photons in a confined optical driving field, which is generated along a slit between two coupled CdS nanowires. **b,** Normalized $z$-axis Poynting vector of an optical driving field, generated with a slit size of 20 nm, a nanowire diameter of 500 nm, and a wavelength of 1.8 μm. The close-up image (upper panel) shows the extremely confined field with an FWHM$_S$ of 77 nm. **c,** $I(t)$ of the virtual ICS pulse, giving a FWHM$_T$ of 19 as (upper panel). For reference, photon numbers $n(t)$ is also provided (orange circle), which basically overlaps with $I(t)$. For comparison, $I(t)$ of a virtual ICS pulse generated in a Gaussian-profile optical driving field is also provided (red dotted line, lower panel). DSC: deep-sub-cycle pulse, Gaussian: Gaussian-profile pulse. **d,** $E_y(t)$ of the DSC pulse in (**c**). **e,** Fourier spectrum $E(\nu)$ of the pulse in (**d**)**,** giving a peak frequency of ~1.9 PHz.

When the electron passes through the driving field along the $z$-direction (Fig. 2b), due to the negligible energy loss of the electron compared with its initial energy (i.e., 1 MeV), $V_e$ can be assumed as a constant (i.e., 0.94$c$) and the trajectory of the electron is almost a straight line (see also section 1.2.4 in Supplemental Material for details). Meanwhile, since the electron-photon-interaction area (i.e., the spot for generating the ICS pulse) is comparable to the wavelength of the ICS photons, the ICS pulse will have a spherical wavefront when it propagates to the spatial far field. Thus we can assume a spherical receiving surface centered on the center of the slit (i.e., at $y$=$z$=0, see also Fig. S1), or equivalently, a concave mirror group to guide the scattered photons out and focused it back into a point in the far field with an almost equal optical path for photons generated with different $\theta$ (Fig. 2a). In the laboratory coordinate system, the angular distribution of the scattered photons $n(z)$ can be described by a factor $f(\theta)$, and the number of scattered photons received at time $t$ and angle $\theta$ can be approximately obtained as[22]

$$n(t,\theta) = n(z) \cdot f(\theta), \tag{3}$$

where the path-dependence (i.e., $z$-dependence) is converted to the time-dependence (i.e., $t$-dependence) by $t=z\cdot(1-\beta\cos\theta)/V_e$, and $f(\theta)=K \cdot \chi^2(1+2\chi(\chi-1))$[22], in which K is a constant coefficient and $\chi=1/(1+\theta^2/(1+\beta^2))$.

Correspondingly, the intensity of the ICS pulse at time $t$ and angle $\theta$ can be obtained as

$$I(t,\theta) = n(t,\theta) \cdot h\nu\,, \tag{4}$$

where $\nu = \nu_0/(1-\beta\cos\theta)$.

By integrating over all possible $\theta$, we obtain the temporal profile of the virtual pulse (i.e., $I$(t)) with a temporal full width at half maximum ($FWHM_T$, i.e., $\tau_1$) of 19 as (Fig. 2c), which is much narrower than that of an ICS pulse generated with the same electron but a 1.8-μm-wavelength diffraction-limited driving field with a Gaussian-profile intensity distribution (red dotted line in Fig. 2c).

Since the phase of the scattered field is inherited from the optical driving field[32], to maximize the peak intensity of the ICS pulse, we assume that the amplitude of the driving field reaches the positive maximum when the electron arrives at the central point of the slit (i.e., $z$=0). Thus, in the spatial far field, the transverse scattered electric field $E_y(t)$ can be obtained as (see also section 1.2.3 in Supplemental Material for details)

$$E_y(t,\theta) = \sqrt{2\sqrt{\frac{\mu_0}{\varepsilon_0}} \cdot I(t,\theta) e^{i2\pi\nu_0\left(\frac{t}{1-\beta cos\theta}\right)}}\,, \tag{5}$$

where $\varepsilon_0$ and $\mu_0$ are the vacuum permittivity and the magnetic permeability, respectively. By integrating over all possible $\theta$, we obtain $E_y(t)$ shown in Fig. 2d, which agrees well with the classical radiation model of an arbitrarily accelerated electron (see also section 1.2.5 in Supplemental Material for details). Unlike the conventional dipolar pulses, here the central peak of the pulse with dominant amplitude is generated by the unidirectional acceleration of the electron driven by the optical driving field, and is thus a unipolar pulse with non-zero electric area [33-36]. Figure. 2e gives the Fourier spectrum of such a virtual pulse, revealing a peak frequency $\nu_p$ of 1.9 PHz (corresponding to a $T_{c1}$ ~0.52 fs). After the peak frequency $\nu_p$, the amplitude of $E(\nu)$ decreases quickly with increasing frequency. As $FWHM_T$ (19 as) is much shorter than $T_{c1}/2$ (i.e., 0.26 fs), such an ICS-generated pulse is a deep-sub-cycle pulse.

It is worth mentioning that, compared with the temporal profiles of previously reported ultrafast sub-cycle or multiple-cycle pulses[14,16,22], here the deep-sub-cycle pulse has a spike-like profile much sharper at the center, which enables a much tighter temporal confinement. For ease of comparison, we define a temporal confinement factor $\zeta=T_{c1}/\tau_1$, in which $\zeta>2$ sets a deep-sub-cycle criterion that breaks the conventional half-cycle limit. For example, for the virtual pulse mentioned above, the calculated $\zeta$ ~27.

**Deep-sub-cycle optical pulses obtained with electron-bunch incidence**

Using single-electron incidence (i.e., $N_e$=1), the photon number of a single ICS pulse ($N_{SP}$) is too low (<<1) to compose a real pulse with $N_{SP}$>>1. For reference, when using a power density of the optical driving field approaching the damage threshold of the CdS crystal (~$1.7\times10^{16}$ W/m$^2$ [37,38]), the calculated $N_{SP}$ ~ $2.5\times10^{-8}$. To increase $N_{SP}$, we propose to use a high-density electron bunch[39-43] (i.e., to increase the electron number), optical material with a higher optical damage threshold to construct the CNP (i.e., to increase the photon number in the optical driving field), driving field with a longer wavelength and thus a larger slit size (i.e., increase the ICS volume and thus the electron and photon numbers involved). Meanwhile, to maintain a short transition time $\delta_t$ of the electron in the enlarged slit size, we use an electron bunch with higher electron energy. Specifically, we use the following parameters: (1) an electron bunch with a duration of 50 as (satisfying a Gaussian distribution), an electron charge of 5 pC (corresponding to an electron density of $1.3\times10^{28}$/m$^3$), an electron energy of 2 MeV, and an elliptical cross-section (200 nm and 1 μm for the short and the long axes, respectively); (2) a pulsed driving field with a central wavelength of 4.4 μm and a pulse width of 100 fs [44]; (3) a CNP consisted of two identical sapphire hexagonal prisms (Fig. 3a), with a diagonal diameter of 1.2 μm, a slit size of 200 nm (corresponding to an FWHM$_S$ of 808 nm), and a power density of $1.3\times10^{18}$ W/m$^2$ at the center of the slit (see Fig. S8 in section 2 of Supplementary Materials) that keeps the power density at the inner edge of the slit below the optical damage threshold of $2\times10^{18}$ W/m$^2$ of the sapphire[45]. In addition, similar to the single-electron-incidence case, the energy loss of the focused electron bunch is negligible during the ICS process.

With the above-mentioned assumptions, we obtained an ICS pulse shown in Fig. 3b, which is composed of a series of single-electron-incidence sub-pulses (see Fig. S4 in section 1.2.2 of Supplemental Material). Owing to the significantly increased electron number (from $N_e$=1 to $3.1\times10^7$) and driving-field photon number, the $N_{SP}$ of the ICS pulse is increased to ~270. Meanwhile, despite of the temporal broadening by the pulse width of the electron bunch (i.e., $\tau_e$=50 as), the overall ICS-pulse width $\tau_I$ (i.e., FWHM$_T$) of 99 as, remains much narrower than that of an ICS pulse generated with the same electron bunch but a 4.4-μm-wavelength diffraction-limited driving field with a Gaussian-profile intensity distribution (red dotted line in Fig.3b). Figure. 3c gives calculated $E_y(t)$ of the ICS pulse shown in Fig. 3b, which retains the unipolar feature of the single-electron-incidence ICS pulse (Fig. 2d). Figure. 3d gives $E(v)$, the Fourier transform of the $E_y(t)$ in Fig. 3c, revealing a peak frequency $v_p$ of 2.6 PHz (corresponding to a $T_{c1}$ ~0.38 fs), which is much higher than that in Fig. 2e (i.e., 1.9 PHz) due to the higher electron energy. Despite the broadening of $\tau_I$, calculated $\zeta$($T_{c1}/\tau_I$=3.8) remains much larger than 2.

By changing the wavelength (i.e., $\lambda_0$) and/or the confinement (i.e., FWHM$_S$) of the optical driving field, it is possible to change $\zeta$ within a wide range. Figure. 3e gives the dependence of $\zeta$ on $\lambda_0$ and FWHM$_S$ of the driving field, with the same 2-MeV electron bunch used in Fig. 3b. It shows that, when $\lambda_0$ is increased from 1 to 5 μm, $\zeta$ can be increased from ~ 1 to larger than 5. With the same $\lambda_0$, $\zeta$ can be increased by reducing FWHM$_S$. However, with increasing $\zeta$, $N_{SP}$ decreases (Fig. 3f). For reference, when $\zeta$ is increased from 4.0 to 4.2, $N_{SP}$ decreases quickly from 270 to 130. On the other hand, using a decreased $\zeta$ (e.g., using a larger slit width and thus a larger FWHM$_S$), $N_{SP}$ can be significantly increased. For example, when $\zeta$ is decreased to 3.2 (still larger than 2), $N_{SP}$ can be increased to 1260.

Therefore, there's a trade-off between $\zeta$ and $N_{SP}$.

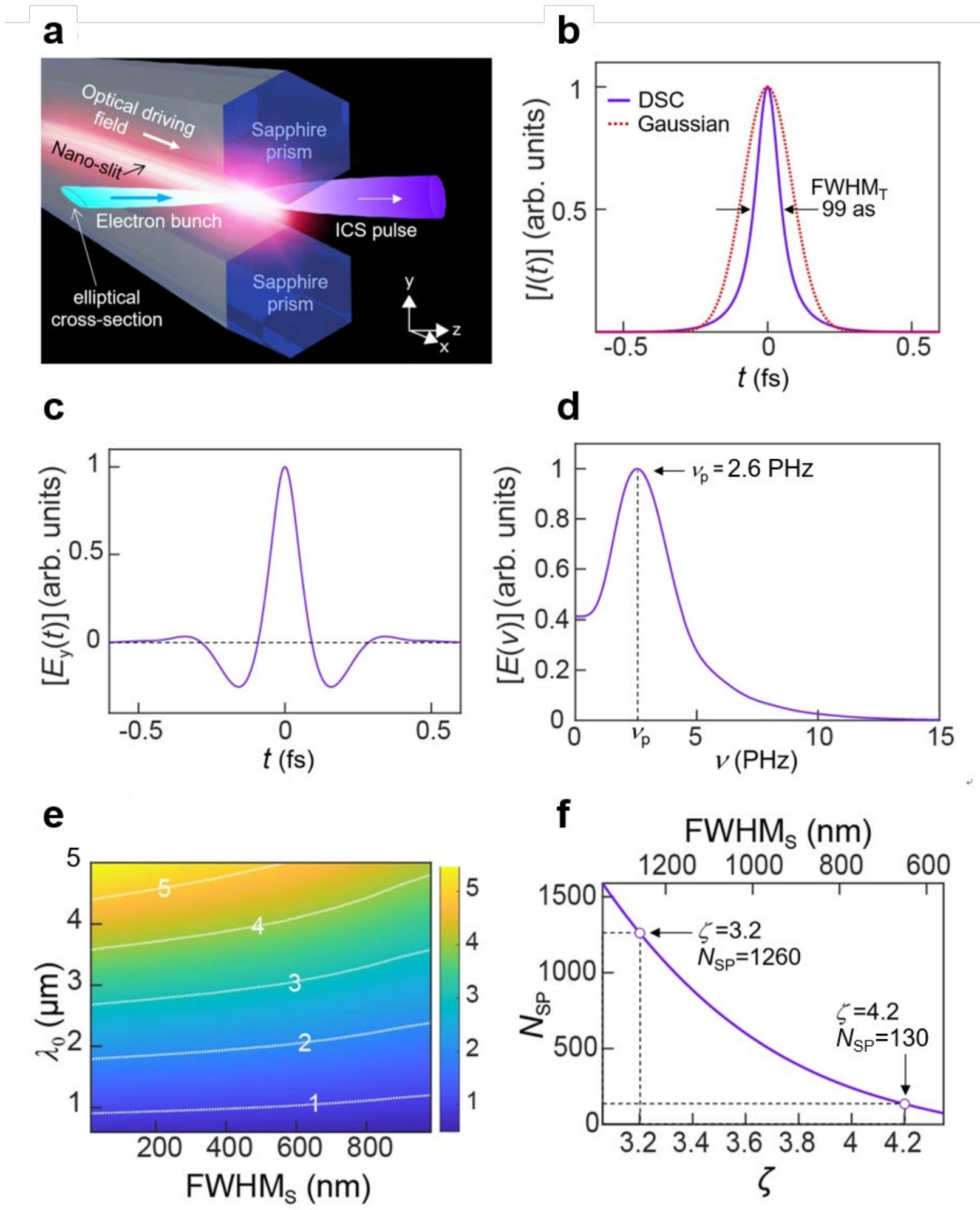

**Fig. 3| Generation of deep-sub-cycle attosecond pulses with attosecond electron bunch and mid-infrared optical driving fields. a,** Schematic diagram of the ICS system. A focused electron bunch with an elliptical cross-section is incident from the left side of a slit, collides with photons in the confined optical driving field that is generated along the slit between two coupled sapphire prisms. **b,** $I(t)$ of the ICS pulse, generated via ICS of a 50-as 2-MeV electron bunch in a pulsed driving field with a $FWHM_S$ of 808 nm, a wavelength of 4.4 μm and a pulse width of 100 fs. The measured pulse width ($\tau_1$) is 99 as. For comparison, $I(t)$ of an ICS pulse generated in a 4.4-μm-wavelength Gaussian-profile optical driving field is also provided (red dotted line). DSC: deep-sub-cycle pulse, Gaussian: Gaussian-profile pulse. **c,** $E_y(t)$ of the DSC pulse in (**b**). **d,** Fourier spectrum $E(\nu)$ of the pulse in (**c**), giving a peak frequency of ~2.6 PHz. **e,** Dependence of $\zeta$ on $FWHM_S$ and $\lambda_0$ of the optical driving field generated in two coupled sapphire prisms in Fig. 3, with a 50-as 2-MeV electron bunch. **f,** Dependence of $N_{SP}$ on $\zeta$ and $FWHM_S$, with an optical driving field of 4.4-μm wavelength and a $2\times10^{18}$ W/m$^2$ peak power density, and a 50-as 2-MeV 5-pC electron bunch.

To significantly increase the $N_{SP}$ while maintaining a relatively large $\zeta$ for practical applications, we propose to use a deep-subwavelength confined THz driving field with a much larger slit width to allow much larger numbers of photons and electrons involved in the ICS process, relying on recently available high-intensity ultrafast THz pulses [46,47]. Figures. 4a-c give $I(t)$, $E(t)$ and $E(\nu)$ of an ICS pulse generated using a 3-MeV 1-fs 1-nC electron bunch and a 0.4-THz ($\lambda_0$=750 μm) 1-ps pulsed driving field with a slit width of 25 μm (corresponding to a $FWHM_S$ of 62.8 μm). Here $\tau_1$ is ~3.6 fs, much narrower than the 19-fs width of a Gaussian profile (Fig. 4a). Compared to that in Fig. 3c, the profile of $E(t)$ has a higher unipolarity (Fig. 4b), due to a larger $\lambda_0$-to-$FWHM_S$ ratio (12 versus 5.4). Calculated $\nu_p$ is ~26 THz (Fig. 4c), corresponding to a $T_{c1}$ of ~38 fs (i.e., $\lambda_1$~12 μm) and consequently a $\zeta$ of ~11.

Especially, owing to the significantly increased numbers of photons and electrons in the ICS process, even with a power density of 3.8×10$^{15}$ W/m$^2$ (far below the optical damage threshold of the sapphire) at the slit center (obtained by assuming a 5% efficiency of coupling a 1.4-mJ 0.4-THz 1-ps pulse [46] into the driving field, a 20-fold underestimation of coupling efficiency compared with that in the near-infrared range [27]), the calculated $N_{SP}$ increases to ~2.3×10$^6$. $N_{SP}$ may be further increased by increasing the peak power density of the driving field and/or the electron density of the electron bunch.

The above-mentioned $\nu_p$ of the ICS pulse is in the mid-infrared region, it can be shifted to near-infrared, visible and even UV spectral ranges by using electron bunch with higher energy and shorter pulse width. Figure. 4d gives $I(t)$ of a near-infrared femtosecond ICS pulse obtained using a 10-MeV 1-fs 1-nC electron bunch and the same 0.4-THz driving field. Benefitting from the increased electron energy, $\nu_p$ increases to ~0.25 PHz (corresponding to a wavelength of ~1.2 μm), with $\tau_1$~1.3 fs, $\zeta$~3.0 and $N_{SP}$~2.3×10$^6$. By further increasing the electron energy and decreasing the pulse width of the electron bunch, we obtain deep-sub-cycle attosecond pulses in the visible or UV spectral range. Figure. 4e gives $I(t)$ of a visible attosecond pulse obtained using a 15-MeV 50-as 50-pC electron bunch and the same 0.4-THz driving field, offering a $\nu_p$ of ~0.49 PHz (corresponding to a wavelength of ~620 nm), a $\tau_1$ of ~180 as, a $\zeta$ of ~11 and a $N_{SP}$~1.1×10$^5$. Figure. 4f gives $I(t)$ of a UV attosecond pulse obtained using a 30-MeV 50-as 50-pC electron bunch and the same 0.4-THz driving field, with $\nu_p$ ~2.0 PHz (corresponding to wavelength of ~150 nm), $\tau_1$~86 as, $\zeta$~5.8 and $N_{SP}$~1.1×10$^5$. Compared to the pulse obtained using a mid-infrared driving field in Figs. 3b-d, $N_{SP}$ can be increased by about 2 to 4 orders of magnitude when using a THz driving field.

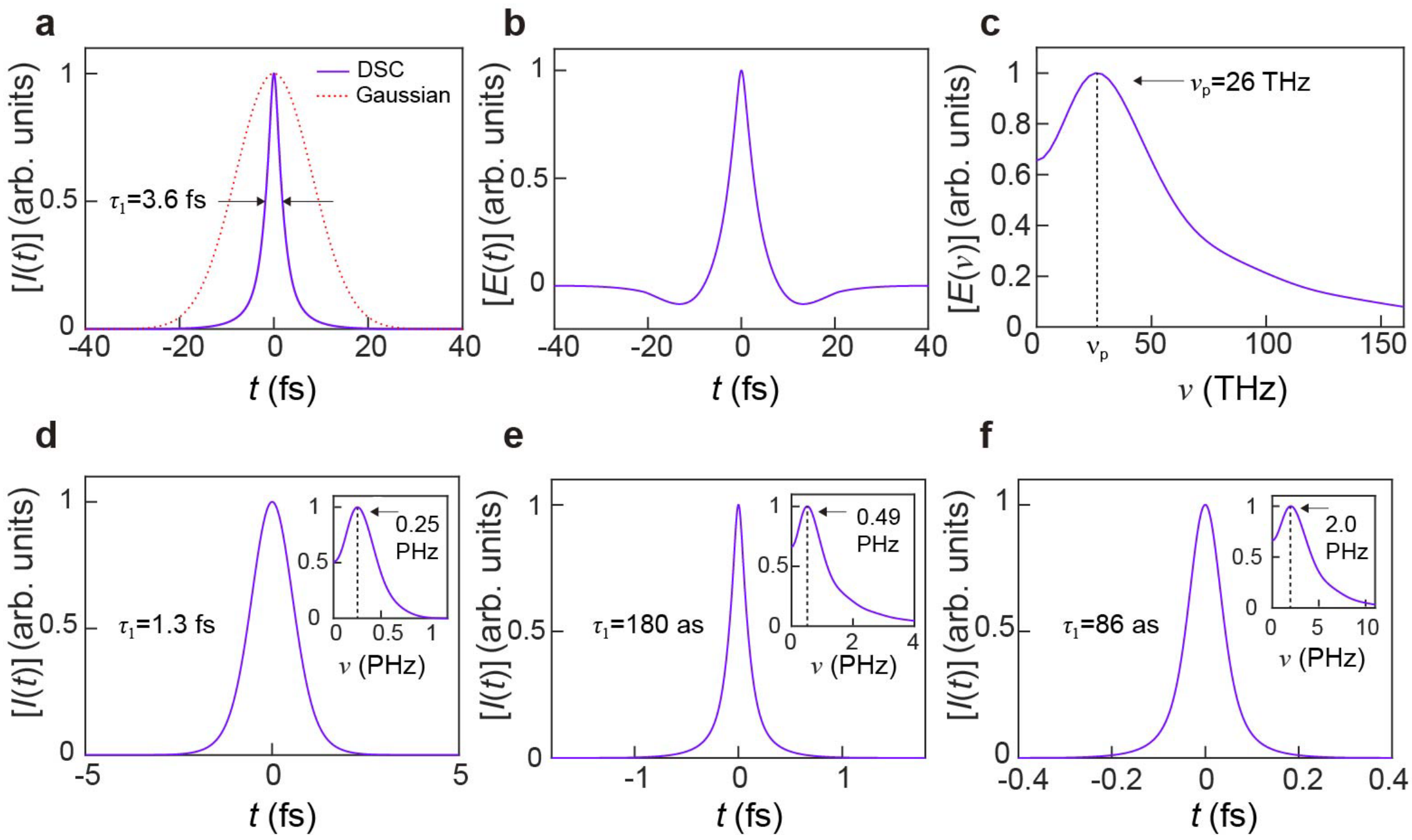

**Fig. 4| Generation of deep-sub-cycle femtosecond and attosecond pulses with ultrafast electron bunches and THz driving optical fields.** **a,** $I(t)$ of an ICS pulse, generated with a 1-fs 3-MeV electron bunch and a 0.4-THz 1-ps pulsed driving field with an FWHMs of 62.8 um. The measured $\tau_1$ is ~3.6 fs. For comparison, $I(t)$ of an Gaussian-profile ICS pulse is also plotted (red dotted line). DSC: deep-sub-cycle pulse, Gaussian: Gaussian-profile pulse. **b,** $E(t)$ of the DSC pulse in (**a**). **c,** Fourier spectrum $E(\nu)$ of the pulse in (**b**), giving a peak frequency $\nu_p$ of ~26 THz. **d,** $I(t)$ of an ICS pulse generated with a 1-fs 10-MeV electron bunch, offering a $\tau_1$ of ~1.3 fs. Inset: $E(\nu)$ with a $\nu_p$ of ~0.25 PHz. **e,** $I(t)$ of an ICS pulse generated with a 50-as 15-MeV electron bunch, offering a $\tau_1$ of ~180 as. Inset: $E(\nu)$ with a $\nu_p$ of ~0.49 PHz. **f,** $I(t)$ of an ICS pulse generated with a 50-as 30-MeV electron bunch, offering a $\tau_1$ of ~86 as. Inset: $E(\nu)$ with a $\nu_p$ of ~2.0 PHz. Driving fields used in (**d**)-(**f**) are all the same used in (**a**).

### Temporal evolution of a deep-sub-cycle pulse propagating in free space

After its generation, the ICS pulse will propagate from the nano-slit region to free-space far field with a spherical wavefront around the *z*-axis within a certain solid angle (Fig. 5a), followed by a slightly slower and much less divergent electron bunch at a small deflected angle with respect to the *z*-axis (see also Fig. S6 in Supplemental Material for details). To investigate its temporal evolution, we calculate the $E_y(t)$ and carrier envelope (CE) of an ICS pulse generated with a 0.4-THz driving field and received on a spherical surface after different propagation distances (i.e., different *z*), with typical results shown in Figs. 5b-d. When the receiving surface is relatively close to the nano-slit center, e.g., $z$=1 cm, $E_y(t)$ and CE has an obvious asymmetric profile (Fig. 5b), due to the nonnegligible amplitude of the asymmetric electrostatic field of the electron bunch, which changes sign when passing through the nano-slit center at $z$=0. When the receiving surface is placed relatively far away from the nano-slit center, e.g., $z$=10 cm, compared to that of the ICS field (decays at a rate of $1/z$), the amplitude of the electrostatic field (decays at a rate of $1/z^2$, see also Fig. S10) reduces to a negligible level, $E_y(t)$ and CE becomes more symmetric (Fig. 5c). When we further increase the distance to $z$=100 cm, the electrostatic field is completely negligible, $E_y(t)$ and CE have stable symmetric profiles (Fig. 5d), showing a stable CE of the deep-sub-cycle ICS pulse propagating in far-field free space.

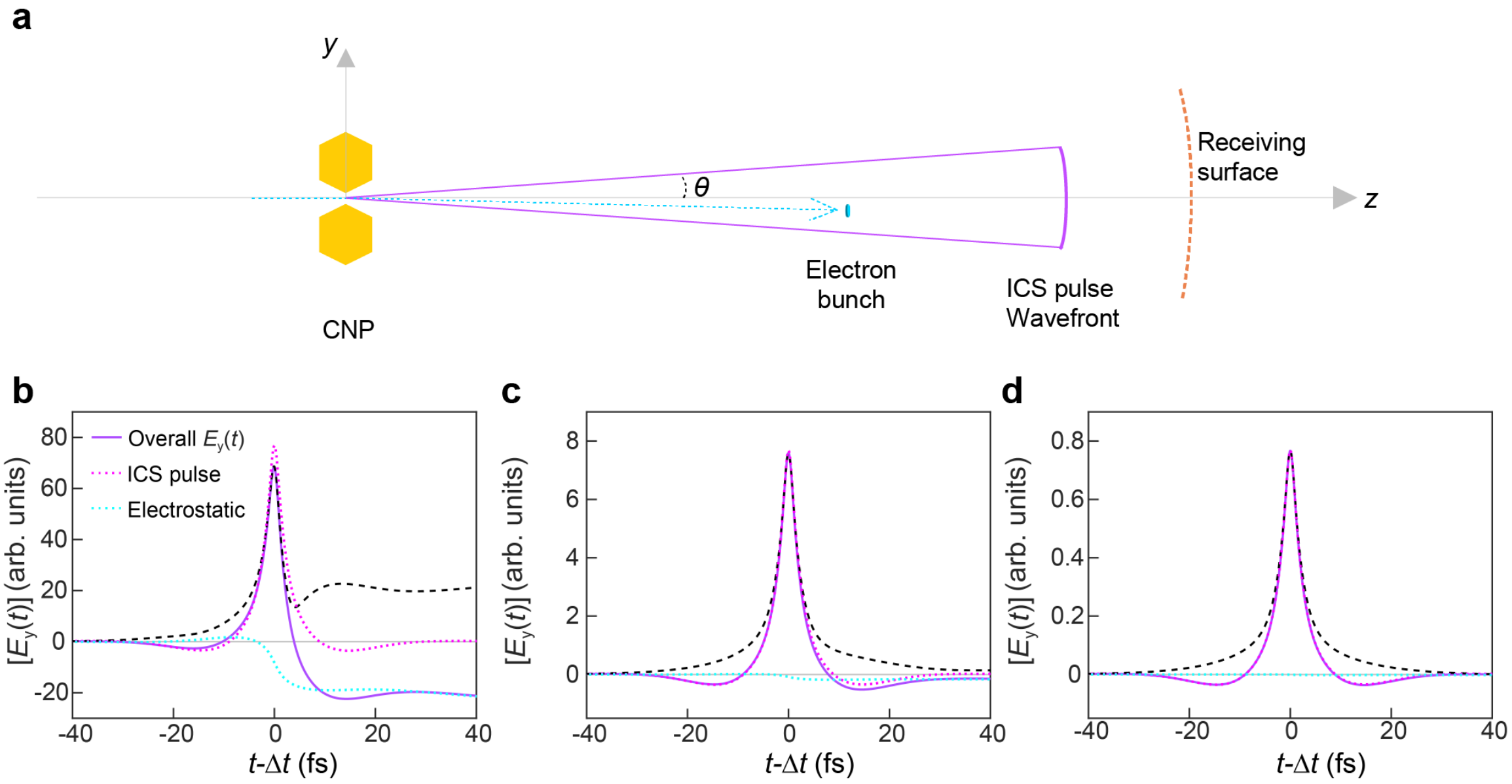

**Fig. 5| Temporal electric field evolution of an ICS pulse. a,** Schematic illustration of an ICS pulse propagating from the driving-field area (around the CNP) to far-field free space. At the far field, the ICS pulse has a spherical wavefront (purple arc) propagating at light speed, followed by a slower and less divergent electron bunch (cyan ellipsoid) with a small deflected angle. The spherical receiving surface (orange dashed arc) is displaced along the z-axis to change the propagation distance of the ICS pulse. **b-d,** Temporal evolution and CE of the ICS pulse during propagation. Here $E_y(t)$ is the calculated overall transverse electric field (purple solid line) combining both the ICS pulse (magenta dotted line) and the fluctuation of the electrostatic field of the electron in the *y*-axis (cyan dotted line), with a position of the center of the receiving surface *z* assumed to be (**b**) $z$=1 cm, (**c**) $z$=10 cm, and (**d**) $z$=100 cm for a 0.4-THz driving field. The carrier envelope of $E_y(t)$ is also given (black dashed line). $\Delta t=\Delta z/c$ is the pulse propagation time from the nano-slit center (i.e., $z$=0) to the receiving surface.

### Conclusion

So far we have demonstrated an approach to generating a deep-sub-cycle attosecond or femtosecond pulse based on ICS from relativistic electrons in a deep-sub-wavelength confined optical driving field. Compared with existing methods for sub-cycle ultrafast pulse generation (e.g., OPCPA, OAWG, and RES) that have not yet reached the half-cycle limit, our approach shown here can break the half-cycle limit and achieve deep-sub-cycle pulses with $\zeta$ >2 (see section 3 in Supplemental Material), while maintaining a stable CE when propagating in the far-field free space. Also, the single-pulse photon number of the deep-sub-cycle can be further increased when higher-intensity THz pulse and/or higher-density relativistic electron bunch are available. As dipolar/multipolar oscillation in optical polarization or resonance is, in general, the underlying mechanism of light-matter interaction[48], and the half-cycle pulse width is the temporal resolution limit of the current ultrafast optical technology [35,49], the deep-sub-cycle optical pulse demonstrated here can not only break the half-cycle limit and reveal deep-sub-cycle dynamics in light-matter interaction, but also manipulate (e.g., excite or mediate) optical response on the deep-sub-cycle level and generate previously difficult-to-reach or unreachable processes (e.g., non-dipole photoionization[50], anharmonic or temporal dark states) and pave a way to temporal super-resolution optical technology ranging from deep-sub-cycle ultrafast microscopy, spectroscopy to atom/molecule manipulation.

**Data and materials availability**

All data are available in the main text or the supplementary materials.

**Acknowledgements** We thank Y. R. Shen for valuable discussions. This work was supported by New Cornerstone Science Foundation (NCI202216) and National Natural Science Foundation of China (62175213, 92150302).

**Author contributions** L.T. conceived of and supervised the project. H.D. and J.G. performed the calculations. All authors analyzed the data. L.T., H.D., and J.G. wrote the manuscript. All authors discussed the results and commented on the paper.

**Competing interests** The authors declare no competing interests.

# Supplemental Material

## Deep-sub-cycle ultrafast optical pulses

**Hongliang Dang,[+] Jiaxin Gao,[+] Hao Wu, Xin Guo & Limin Tong[*]**

*New Cornerstone Science Laboratory, State Key Laboratory of Extreme Photonics and Instrumentation, College of Optical Science and Engineering, Zhejiang University, Hangzhou 310027, China*

[+] These authors contribute equally.

* Corresponding author: phytong@zju.edu.cn

## Notations

This work involves the transformation of variables between two coordinate systems — the E. R. frame (electros at rest frame) and the Lab frame (i.e., the stationary observer frame). We denote that in the E. R. frame with a superscript '′', and that in the Lab frame without superscript.

Table 1 gives a list of notations involved in this work.

**Table 1.** Notations

| | |
|---|---|
| **Coordinate** | |
| $\phi$ | incident angle in the Lab frame |
| $\phi'$ | incident angle in the E. R. frame |
| $\theta$ | scattered angle of the ICS photon in the Lab frame |
| $\theta'$ | scattered angle of the ICS photon in the E. R. frame |
| $\Omega$ | solid angle |
| $\psi$ | azimuthal angle projected at the transverse plane (i.e., *xy* plane) |
| $x$ | $x$ coordinate, the direction of the driving optical field propagation |
| $y$ | $y$ coordinate, the direction of the electric field of the driving optical field in the nano-slit |
| $z$ | $z$ coordinate, the direction of the electron incidence |
| $\hat{\boldsymbol{r}}$ | the unit vector pointing from the electron to the observer |
| $r$ | distance between the electron and the observer |
| $t_g$ | the moment when the ICS photon is generated in the Lab frame |
| $t$ | the moment when the ICS photon reaches the observer in the Lab frame |
| **Driving optical field** | |
| $\nu_0$ | optical frequency |
| $T_{c0}$ | oscillation period |
| $\lambda_0$ | wavelength |
| $\tau_0$ | pulse width |
| $N_0$ | photon number |
| $E_{p0}$ | pulse energy |
| $\boldsymbol{E}_{\mathbf{driving}}$ | electric field of the driving optical field |
| $A_0$ | amplitude |
| $A_{0N}$ | normalized amplitude |
| $\varphi$ | phase |
| $g(t_g)$ | $t_g$-dependent envelope |
| $\boldsymbol{B}_{\mathbf{driving}}$ | magnetic field of the driving field |
| $w$ | slit width of the nano-slit structure |
| $P_{\mathrm{mode}}$ | average mode power, i.e., $P_{\mathrm{mode}}= E_{p0}/\tau_0$ |
| $\boldsymbol{S}(\boldsymbol{z},\boldsymbol{y})$ | Poynting vector along the propagation direction |
| $P(z)$ | z-dependent power density distribution, i.e., $P(z)=\int_{-w/2}^{w/2}\boldsymbol{S}(\boldsymbol{z},\boldsymbol{y})dy\,/w$ |
| $\mathrm{FWHM_S}$ | spatial FWHM of the intensity distribution along the z-axis |
| **Incident electron/electron bunch** | |
| $E_e$ | electron energy |
| $\boldsymbol{V}_{\mathbf{e}}$ | electron speed |
| $\boldsymbol{\beta}$ | normalized $\boldsymbol{V}_{\mathbf{e}}$ over $c$, i.e., $\boldsymbol{\beta}=\boldsymbol{V}_{\mathbf{e}}/c$ |
| $\gamma$ | Lorentz factor, i.e., $\gamma=\sqrt{1/(1-\beta^2)}$ |
| $\tau_e$ | pulse width of the electron bunch |
| $N_e$ | electron number |
| **ICS pulse** | |
| $\nu$ | optical frequency |
| $T_{c1}$ | oscillation period |
| $\lambda_1$ | wavelength |
| $N_{SP}$ | photon number |
| $\nu_{peak}$ | peak frequency |
| $T_{peak}$ | oscillation period at the peak frequency |
| $\tau_1/\mathrm{FWHM_T}$ | pulse width/temporal FWHM |
| $n(t,\theta)$ | ICS photon number received at time $t$ and angle $\theta$ |
| $n(t)$ | ICS photon number received at time $t$, i.e., $n(t)=\int n(t,\theta)d\theta$ |
| $I(t,\theta)$ | ICS pulse intensity received at time $t$ and angle $\theta$, i.e., $I(t,\theta)=n(t,\theta)\cdot h\nu$ |
| $I(t)$ | ICS pulse intensity received at time $t$, i.e., $I(t)=\int I(t,\theta)d\theta$ |
| $\boldsymbol{E}(\boldsymbol{t},\boldsymbol{\theta})$ | electric field of the ICS pulse at time $t$ and angle $\theta$ |
| $\boldsymbol{E}(\boldsymbol{t})$ | electric field of the ICS pulse at time $t$, i.e., $\boldsymbol{E}(\boldsymbol{t})=\int \boldsymbol{E}(\boldsymbol{t},\boldsymbol{\theta})d\theta$ |
| $E(\nu)$ | Fourier spectrum of $\boldsymbol{E}(\boldsymbol{t})$ |
| **Constants** | |
| $c$ | speed of light |
| $h$ | Planck constant |
| $\sigma_T$ | Thomson scattering cross-section |
| $e$ | elementary charge |
| $m_0$ | electron mass |
| $\varepsilon_0$ | vacuum dielectric constant |
| $\mu_0$ | vacuum permeability |
| **Degree of temporal confinement** | |
| $\zeta$ | temporal confinement factor |

The remaining unlabeled notations will be further explained in the text.

# 1. Inverse Compton Scattering (ICS) of electrons in an optical nanoslit mode

## 1.1 General description of an ICS process

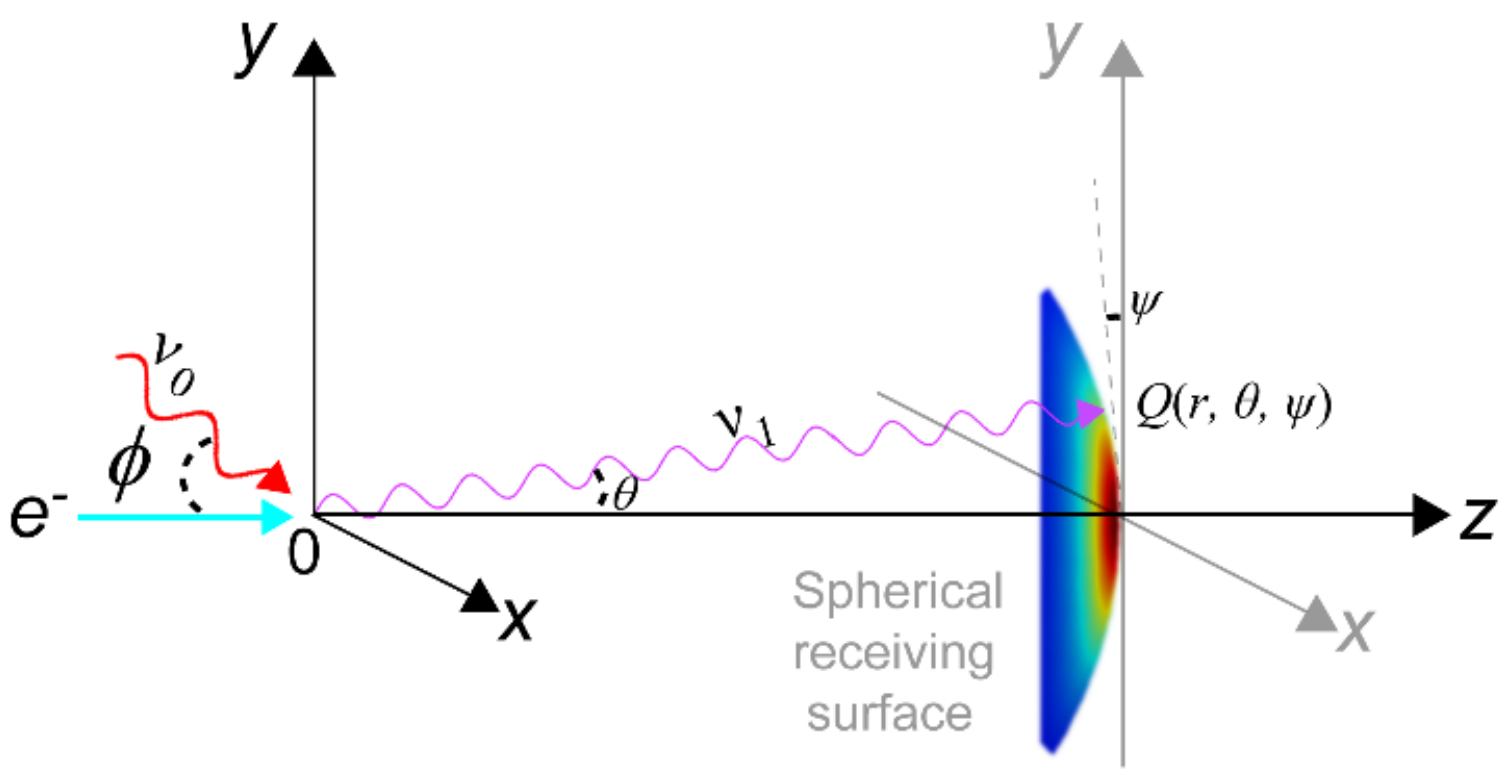

**Fig. S1| Schematic illustration of the ICS system.** Inverse Compton scattering of a photon (red arrow line) and an electron (cyan arrow line) colliding with each other on their motion trajectories. The scattered photon (purple arrow line) will be collected at a spherical receiving surface centered around the original point of $x$=$y$=$z$=0, with an angle $\theta$.

As shown in Fig. S1, for an electron (energy of $E_e$, speed of $V_e$) moving along the $z$-axis, assuming that the incident angle and the scattering angle of a photon (energy of $\nu_0$ and wavelength of $\lambda_0$) are $\phi$ and $\theta$ in the Lab frame, and $\phi'$ and $\theta'$ in the E. R. frame, respectively. Then, in the Lab frame, we have[1,2]

**(1) Frequency of scattered photons**

$$\nu(\phi,\theta) = \nu_0 \frac{1-\beta\cos\phi}{1-\beta\cos\theta} , \tag{1}$$

where $\beta=V_e/c$. In this work, $\phi$=90°, therefore

$$\nu(\theta) = \frac{\nu_0}{1-\beta\cos\theta} \tag{2}$$

and therefore, the wavelength of the scattered photons $\lambda_1=\lambda_0\cdot(1-\beta\cos\theta)$.

**(2) The range of the scattering angle**

$$\sin\theta \in [-\frac{1}{\gamma},\frac{1}{\gamma}] , \tag{3}$$

where the Lorentz factor $\gamma=[1/(1-\beta^2)]^{1/2}$.

**(3) Angle distribution of scattered photons**

The angular distribution of scattered photons can be derived from ref.[3]

$$f(\theta,\psi) = K\chi^2\left[\frac{1}{2}+\chi(\chi-1)(1+cos2\psi)\right] , \tag{4}$$

where $K$ is a constant coefficient, $\chi=1/[1+\gamma^2\theta^2]$ and $\psi$ is the azimuthal angle at the observing plane (i.e., the receiving surface).

**(4) Time compression effect**

Time compression effect is well-known in e.g., synchrotron radiation[4] for an electron moving with relativistic velocity towards the observer. Since the electron and its radiation travel with comparable velocities, the radiation field generated by the electron over a relatively long time ($t_g$=$z$/$V_e$) is received by the observer within a shorter time interval ($t$), with

$$t = t_g(1 - \beta\cos\theta) \quad . \qquad (5)$$

For the ICS process we discussed in this work, we also used relativistic electrons/electron bunches with velocities close to the vacuum light speed, therefore the time compression effect (i.e., Eq.5) should be considered.

## 1.2 ICS pulses generated in an optical nanoslit mode

### 1.2.1 Nano-slit-mode optical driving field

As shown in Fig.S2, in the nano-slit of a coupled nanowire pair (CNP), although the field confinement in the y-direction (~20 nm) is tighter than that in the *z*-direction (~77 nm), to avoid the interaction between the electron and the Cadmium Sulfide (CdS) material when it passing through, here we assume that the electron passes through the nano-slit along the *z*-direction (the white dashed line arrow) and interacts with the optical driving field all the way in the air or vacuum, without hitting the edges of the nano-slit.

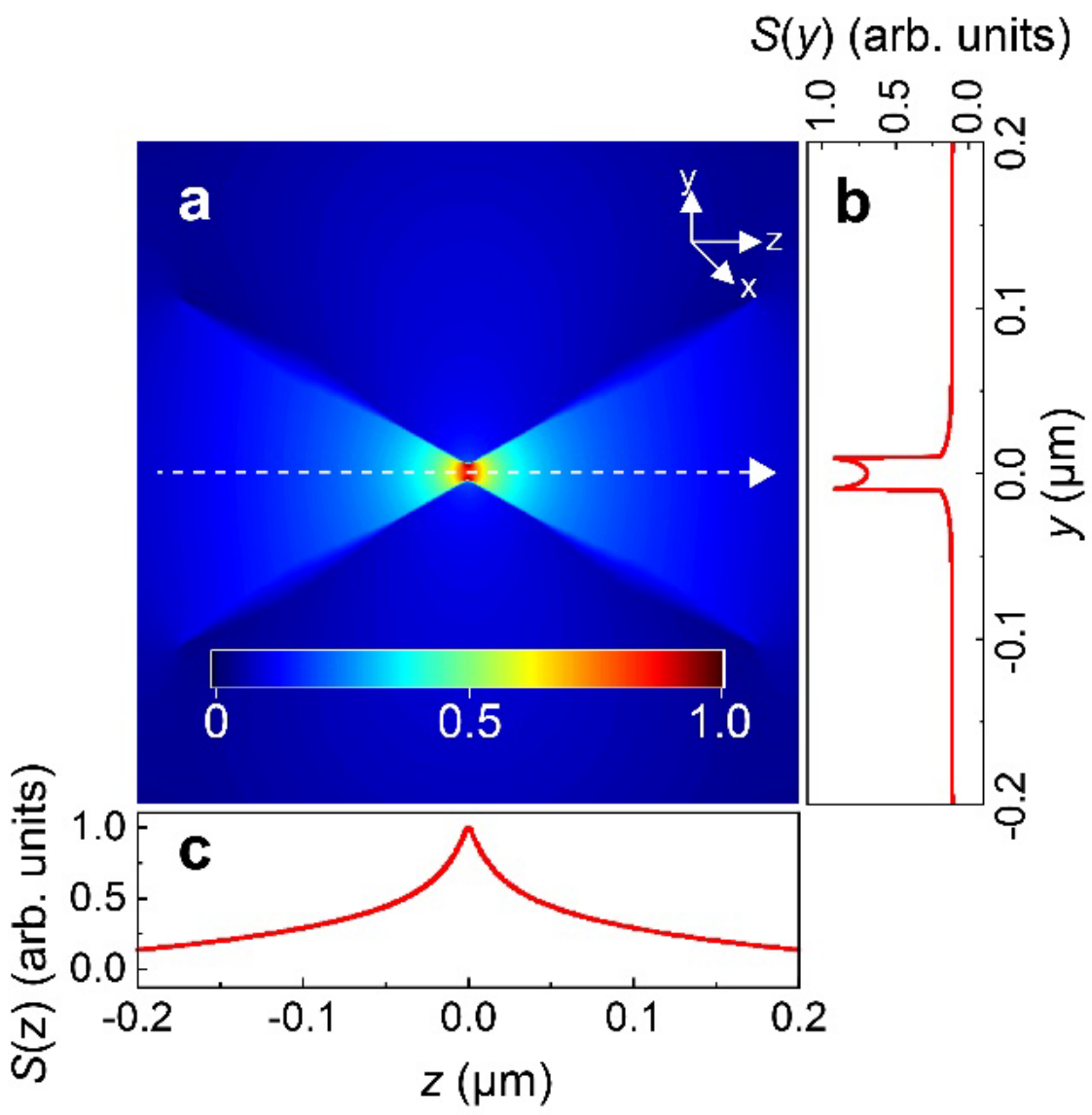

**Fig. S2| Calculated Poynting vector of the nano-slit mode in a CNP constructed with two identical hexagonal CdS nanowires with a diameter of 500 nm, a slit width of 20 nm, and a wavelength of 1.8 μm. a,** 2-dimensional Poynting vector $|\boldsymbol{S}(y,z)|$ in the $y$-$z$ plane. **b** and **c,** 1-dimensional Poynting vector $S(y)=|\boldsymbol{S}(y, z=0)|$ and $S(z) = |\boldsymbol{S}(y=0, z)|$ along $y$- and $z$-axis centrosymmetric lines, respectively. The results are calculated with the finite element method (COMSOL Multiphysics®).

### 1.2.2 Time-dependent photon number and intensity of ICS pulses

Assuming that the optical driving field (i.e., a nano-slit-waveguide mode of a CNP structure[5])

is generated by a femtosecond pulse with a pulse width of $\tau_0$, a pulse energy of $E_{p0}$, and a central frequency of $v_0$, the total number of photons contained in a single-pulse driving field is

$$N_0 = \frac{E_{p0}}{hv_0} \tag{6}$$

where $h$ is the Planck constant.

Assuming the transition time $\delta_t$ of the electron (or electron bunch) passing through the optical driving field is shorter than the pulse width $\tau_0$ (i.e., $\delta_t<\tau_0$), the number of photons involved in the ICS process is $N_0 \cdot \frac{\delta_t}{\tau_0}$ , therefore, the total number of ICS photons is

$$N_{\mathrm{SP}} = N_0 \cdot 1 \cdot \frac{\sigma_{\mathrm{T}}}{A} \cdot \frac{\delta_t}{\tau_0} \tag{7}$$

where $\sigma_T$ is the Thomson scattering cross-section of the electron (~$6.65\times10^{-29}$ $m^2$), and $A$ is the interaction area of electrons and photons.

As shown in Fig. S1, when an electron passes through a certain point $z$ in the driving field, it undergoes inverse Compton scattering and radiates photons with higher frequency (i.e., $v(\theta) = \frac{v_0}{1-\beta\cos\theta}$). Assume that the number density of photons radiated at point $z$ is $n(z)$ (detailed expressions are given in Eq.10 for single electron and Eq.13 for electron bunch, respectively.), which can be normalized by $N_{\mathrm{SP}} = \int n(z)dz$.

Therefore, the fraction of $n(z)$ that can arrive at a position Q($r$, $\theta$, $\psi$) on the receiving surface is

$$n(z, r, \theta, \psi) = n(z) \cdot f(\theta, \psi). \tag{8}$$

When considering the time compression effect, the exact arrival time $t$ of the photons scattered at the point $z$ can also be derived as

$$t = \frac{z}{V_e} \cdot (1 - \beta\cos\theta) + \frac{r}{\mathrm{c}} \tag{9}$$

where $r$ is the distance between the nano-slit center (i.e., the original point of the coordinate system) and the spherical receiving surface, and we assume that $r>>z$ (i.e., the receiving surface is placed in the far field of the nano-slit mode). From Eq.9, we have $z = \frac{V_e}{1-\beta cos\theta}(t - r/c)$, by substituting it into Eq.8, we can convert $n(z, r, \theta, \psi)$ into $n(t, r, \theta, \psi)$.

By integrating $n(t, r, \theta, \psi)$ over $\psi$, we can get $\theta$-dependent photon number density $n(t, r, \theta)$ received at time $t$ on the receiving surface, and subsequently the time-dependent scattered light intensity $I(t)=\int n(t, r, \theta) \cdot hv(\theta)d\theta$ , from which we can obtain the electric field $E(t)$ under the far-field approximation and $E(v)$ by Fourier transform of $E(t)$ for the ICS pulse.

Following the above-mentioned approach, we calculate the ICS photon number in two situations.

**(1) Single-electron incidence**

Assuming that the average mode power is $P_{\mathrm{mode}}= E_{p0}/\tau_0$, and the Poynting vector at an arbitrary point is $\boldsymbol{S}(z, y)$ ($x$ is not included due to the structural symmetry along the $x$-axis). When the electron passes through the nano-slit along the central axis (i.e., $z$-axis), due to $y$=0, the Poynting vector is only relevant with $z$ component (i.e., $\boldsymbol{S}(z)$). Therefore, the number density of scattered photons at

position $z$ is

$$n(z) = N_0 \cdot \frac{S(z)}{P_{mode}} \cdot \sigma_{\mathrm{T}} \cdot \frac{\delta_t}{\tau_0} \tag{10}$$

According to Eqs.8-9, we then convert $n(z)$ to $n(t, r, \theta, \psi)$. Incorporated with Eq.2, and integrated over $\psi$, we obtained the $\theta$-dependent photon intensity at time $t$

$$I(t, \theta) = \int_0^{2\pi} n(t, r, \theta, \psi) h\nu(\theta) d\psi \tag{11}$$

and $t$-dependent intensity by integrating over $\theta$

$$I(t) = \int_{-\arcsin\frac{1}{\gamma}}^{\arcsin\frac{1}{\gamma}} I(t, \theta) d\theta \tag{12}$$

For reference, using the same parameters as Fig. 3, the calculated $I(t, \theta)$ with $\theta$ of 0°, 2°,5°,8° and 10°, is plotted as color lines; and $I(t)$ by integrating all possible $\theta$ is plotted as the black dotted line in Fig. S3. The measured pulse width $\tau_1$, defined as the temporal full width at half maximum ($\mathrm{FWHM_T}$) of the temporal profile of $I(t)$, is ~66 as.

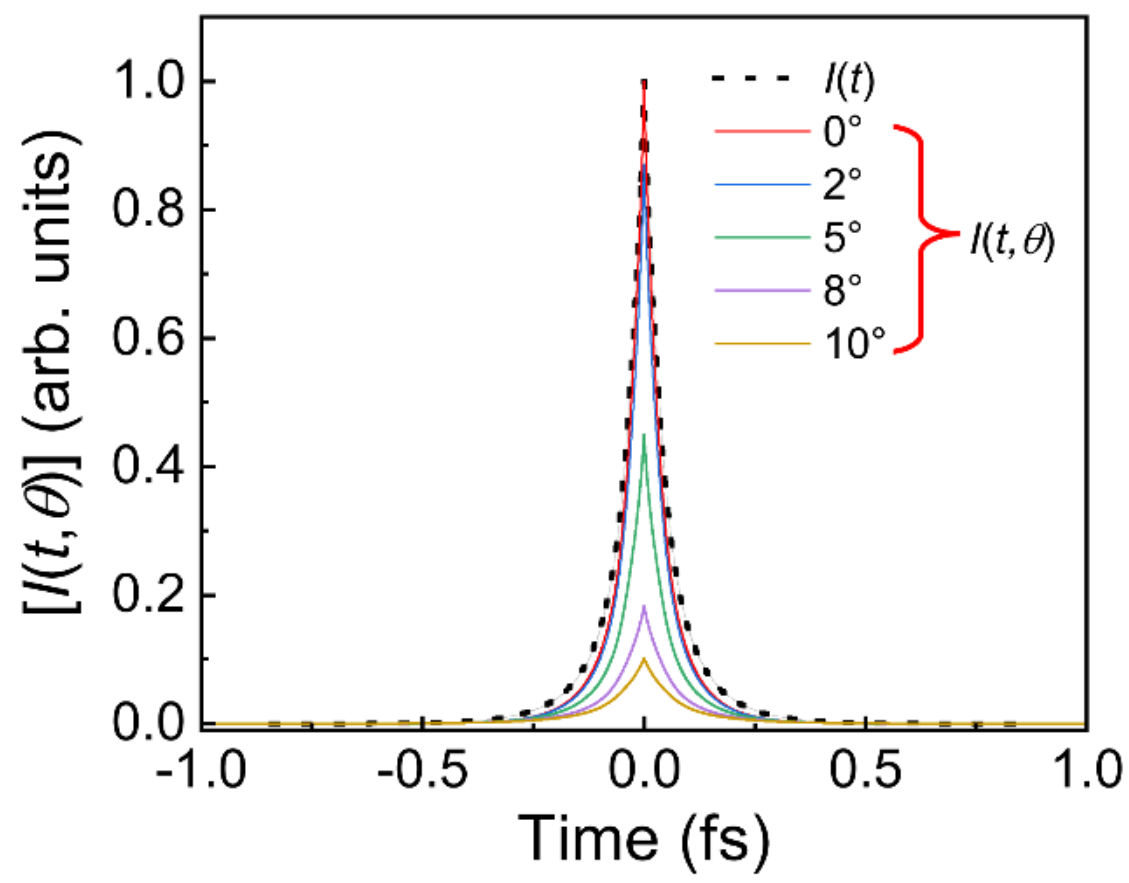

**Fig. S3| Calculated $I(t)$ of the single-electron-incidence ICS pulse.** Waveform $I(t, \theta)$ of the single-electron-incidence ICS pulse received at time $t$, with $\theta$ of 0°, 2°,5°,8°, and 10°, respectively. The overall waveform $I(t)$ is plotted in black dotted line. The results are calculated with the same parameters as in Fig. 3.

**(2) Electron-bunch incidence**

Assuming that the electron bunch has an electron number density of $n_e$ $(x, y, z)$, which can be normalized to the total electron number by $N_e = \iiint n_e dxdydz$. Then, the number density of scattered photons can be obtained as

$$n(z) = N_0 \cdot \frac{\iint n_e\,(x, y, z) \cdot S(x, y, z) dxdy}{P_{\mathrm{mode}}} \cdot \sigma_{\mathrm{T}} \cdot \frac{\delta_t}{\tau_0} \tag{13}$$

In this work, we assume that in an electron bunch with a pulse width of $\tau_e$, the electrons are distributed homogenously along the $y$-axis, and temporally obey a Gaussian distribution, i.e., $n_e = N_{e0} \cdot \exp[-(t_e)^2/\tau_e^2]$ (where $N_{e0} = N_e / \int_{-\infty}^{\infty} e^{-(t_e/\tau_e)^2}$, is a normalized coefficient), in which we assume the moment that the center of the electron bunch passing through the center of the nano-slit (i.e., $z$=0) to be $t_e$=0, and calculate with $t_e$ from -10$\tau_e$ to 10$\tau_e$ (outside this range, $n_e$ is negligible) . Then, we divide the electron bunch into intervals (each has a size of $\delta_{t_e}$) small enough that can be treated as a single electron (i.e., the average electron number in such an interval $\delta_{Ne} = \int_{t_e}^{t_e+\delta_{t_e}} n_e dt_e \sim 1$), and also consider that $S(x,y,z)$ is a constant along the $x$-axis, thus, the number density of scattered photons in an interval $\delta_{t_e}$ is

$$n(z,\delta_{t_e}) = N_0 \cdot \int_{t_e}^{t_e+\delta_{t_e}} n_e dt_e \cdot \frac{\int_{-\frac{w}{2}}^{\frac{w}{2}} S(z,y)dy\,/w}{P_{\text{mode}}} \cdot \sigma_{\text{T}} \cdot \frac{\delta_{\text{t}}}{\tau_0} \tag{14}$$

where $w$ is the slit width of the CNP structure. Similar processes as Eqs.10-12, we can convert $n(z, \delta_{t_e})$ to $n(t, \delta_{t_e})$, integrate over $\psi$, and obtain the time-dependent intensity distribution generated by the electrons within a small interval $\delta_{t_e}$

$$I(t,\delta_{t_e}) = \int_{-\arcsin\frac{1}{\gamma}}^{\arcsin\frac{1}{\gamma}} n(t,\delta_{t_e}) \cdot h\nu(\theta)d\theta \tag{15}$$

For reference, using the same parameters as Fig. 3, the calculated $I(t,\delta_{t_e})$ (with $\delta_{Ne} = 1.0$) is given as colored sub-pulses in Fig. S4.

By integrating $I(t,\delta_{t_e})$ over $\delta_{t_e}$, we have

$$I(t) = \int_{-10\tau_e}^{10\tau_e} I(t,\delta_{t_e})\, d\delta_{t_e} \tag{16}$$

Based on Eq.16, we obtain $I(t)$, as the purple envelope plotted in Fig. S4, from which we obtain the pulse width, i.e., $\text{FWHM}_\text{T}$ of the temporal profile of $I(t)$, of ~99 as.

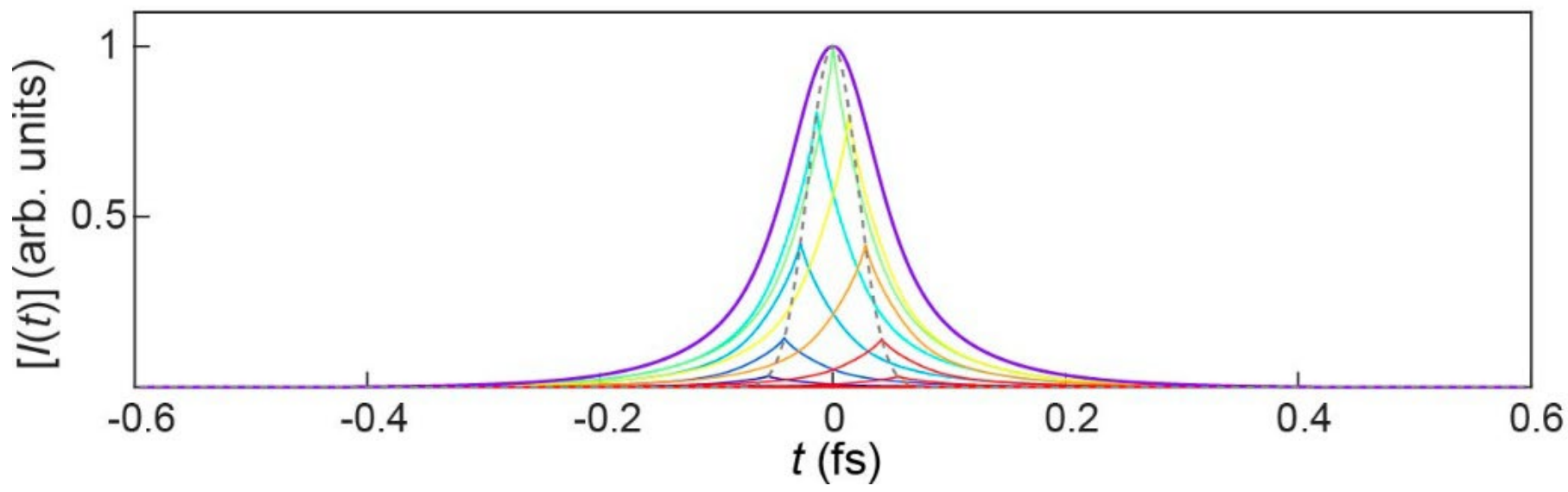

**Fig. S4| Calculated *I*(*t*) of the electron-bunch-incidence ICS pulse.** Waveform $I(t)$ of an overall ICS pulse (purple solid line), which is composed of sub-pulse waveforms $I(t,\delta_{t_e})$ (thin lines with different colors) with a Gaussian-profile temporal distribution (grey dashed line). The results are calculated with the same parameters as in Fig. 3.

### 1.2.3 Electric field *E*(*t*) and Fourier spectrum *E*(*ν*) of the ICS pulse at the receiving surface

Since the phase of the scattered field is inherited from the driving optical field[6], to get a symmetric temporal waveform of the ICS pulse, we assume that the amplitude of the driving field reaches the positive maximum when the electron (or center of the electron bunch) arrives at the central point of the nano-slit (i.e., $z$=0). Thus, in the spatial far field, we obtain the transverse electric field $\boldsymbol{E}_y(\boldsymbol{t},\boldsymbol{\theta})$ of the ICS pulse as

$$\boldsymbol{E}_{\mathrm{y}}(\boldsymbol{t},\boldsymbol{\theta}) = \sqrt{2\sqrt{\frac{\mu_0}{\varepsilon_0}}\cdot I(t,\theta)}e^{i2\pi\nu_0\left(\frac{t}{1-\beta cos\theta}\right)} \tag{17}$$

For reference, Fig. S5 gives calculated $t$ and $\theta$-dependent $\boldsymbol{E}_{\mathrm{y}}(\boldsymbol{t},\boldsymbol{\theta})$ at the receiving surface in the far field.

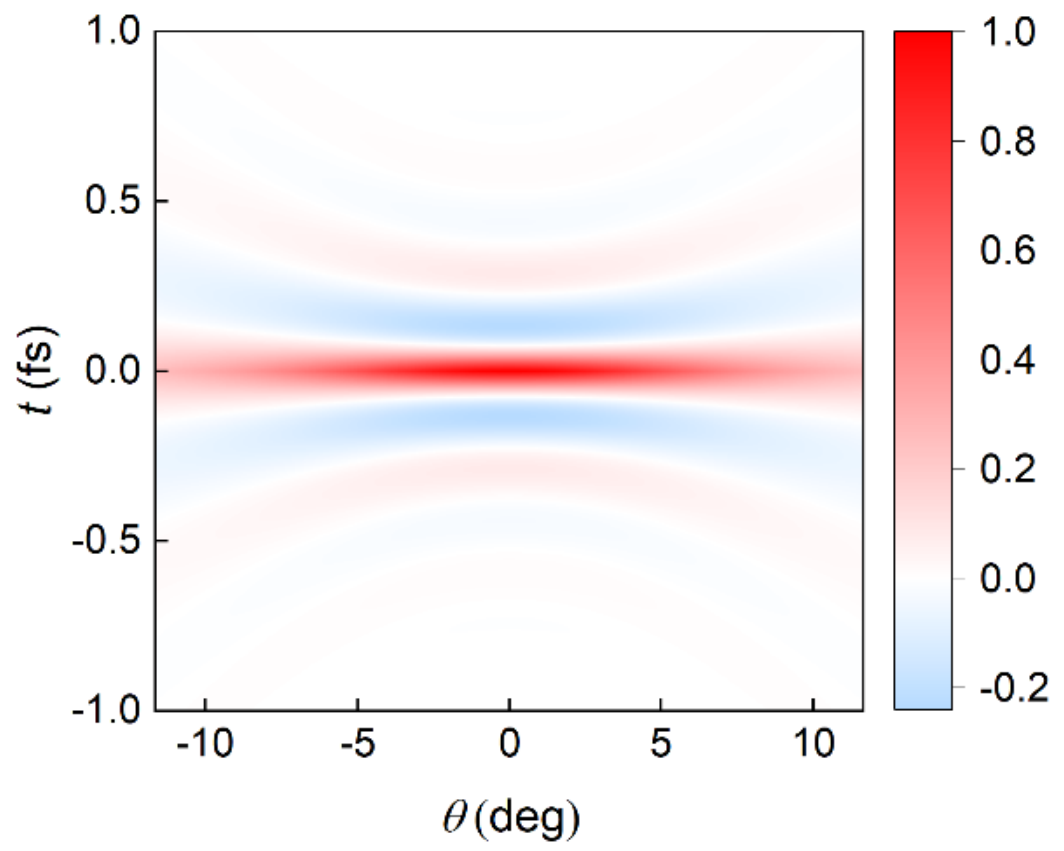

**Fig. S5| Calculated *t* and *θ*-dependent *E*y(*t*,*θ*), at the receiving surface in the far field.** The results are calculated with the same parameters as in Fig. 3

When we integrate $\boldsymbol{E}_{\mathrm{y}}(\boldsymbol{t},\boldsymbol{\theta})$ over $\theta$, we can obtain

$$\boldsymbol{E}_{\mathrm{y}}(\boldsymbol{t}) = \int \boldsymbol{E}_{\mathrm{y}}(\boldsymbol{t},\boldsymbol{\theta})d\theta \tag{18}$$

Then, by performing the Fourier transformation, we can obtain

$$E(\nu) = \int_{-\infty}^{\infty} E(t)e^{-j\cdot 2\pi\nu t}dt \tag{19}$$

## 1.2.4 Electron's trajectory driven by the nano-slit-mode optical driving field

A CNP structure can selectively support a $TE_0$-like nano-slit mode with nearly linear polarization along the $\boldsymbol{y}$ direction[5,7,8], with an electric field of

$$\boldsymbol{E}_{\mathbf{driving}} = A_0 g(t_{\mathrm{g}})cos(\varphi)\hat{\boldsymbol{y}} \tag{20}$$

where $A_0$ is the amplitude of the nano-slit-mode field, $g(t_g)$ is the amplitude envelope and $\varphi$ is the phase of the field (i.e., $\varphi=2\pi\nu_0 t_g$).

Assuming that the incident electron has an initial normalized energy $\gamma$ (i.e., the Lorentz factor), an initial normalized velocity $\boldsymbol{\beta}$, and an initial phase $\varphi$ of 0, then its energy and momentum satisfy the relativistic kinematic equations[9]

$$\frac{d(\gamma\boldsymbol{\beta})}{dt} = -\frac{e}{m_0 c}\left(\boldsymbol{E}_{\mathbf{driving}} + \boldsymbol{V}_{\mathbf{e}} \times \boldsymbol{B}_{\mathbf{driving}}\right) \tag{21}$$

$$\frac{d\gamma}{dt} = -\frac{e}{m_0 c^2}\boldsymbol{V}_{\boldsymbol{e}} \cdot \boldsymbol{E}_{\mathbf{driving}} \tag{22}$$

where $e$ is the elementary charge of the electron, $m_0$ is the rest mass of the electron and $\boldsymbol{B}_{\mathrm{driving}}$

is the magnetic field of the nano-slit-mode optical field. Based on Eq.21 and Eq.22, the electron's velocity can be obtained as

$$\begin{cases} \beta_x = \dfrac{\frac{A_{0N}^2}{2\gamma^2}G_1^2(\varphi)}{1+\frac{A_{0N}^2}{2\gamma^2}G_1^2(\varphi)} \\ \beta_y = \dfrac{A_{0N}G_1(\varphi)}{\gamma\left[1+\frac{A_{0N}^2}{2\gamma^2}G_1^2(\varphi)\right]} \\ \beta_z = \dfrac{\beta}{1+\frac{A_{0N}^2}{2\gamma^2}G_1^2(\varphi)} \end{cases} \tag{23}$$

where $A_{0N}$ is the normalized amplitude of the driving field (i.e., $A_{0N}=\frac{eA_0}{m_0 2\pi\nu_0 c}$), and $G_1(\varphi) = -\int_0^{\varphi} g(\varsigma)cos\varsigma\, d\varsigma$.

From Eq.23, we can obtain the electron's trajectory

$$\begin{cases} x = \frac{c}{2\pi\nu_0}\frac{A_{0N}^2}{2\gamma^2}\cdot\int_0^{\varphi} G_1^2(\varsigma)\, d\varsigma \\ y = \frac{c}{2\pi\nu_0}\frac{A_{0N}}{\gamma}\cdot\int_0^{\varphi} G_1(\varsigma)\, d\varsigma \\ z = \frac{c}{2\pi\nu_0}\beta\varphi \end{cases} \tag{24}$$

Based on Eq.24, we calculate the trajectory of an incident 2-MeV electron passing through a nano-slit of a CNP structure constructed with two identical sapphire hexagonal prisms (with a diagonal diameter of 1.2 μm, a slit width of 200 nm) at 4.4-μm wavelength (the same parameters used in Fig.3), as shown in Figs. S6a-c. After passing through the nano-slit along the *z*-direction for ~10 μm, the electron is slightly deflected in the *y*-direction for about 40 nm (Fig. S6d), a very slight deviation from its original *z*-direction. The deflection of the electron in the *x*-direction is much smaller (~10 fm), and is completely negligible. Therefore, the trajectory of the electron can be largely regarded as a straight line (Figs. S6c-f).

Also, since the transition time of the electron ($\delta_t$) is much shorter than the width of the driving pulse ($\tau_0$), the phase of the driving field is almost unchanged when the electron passes through (Fig. S6a), resulting in a largely unidirectional acceleration and consequently unipolar radiation of the electron.

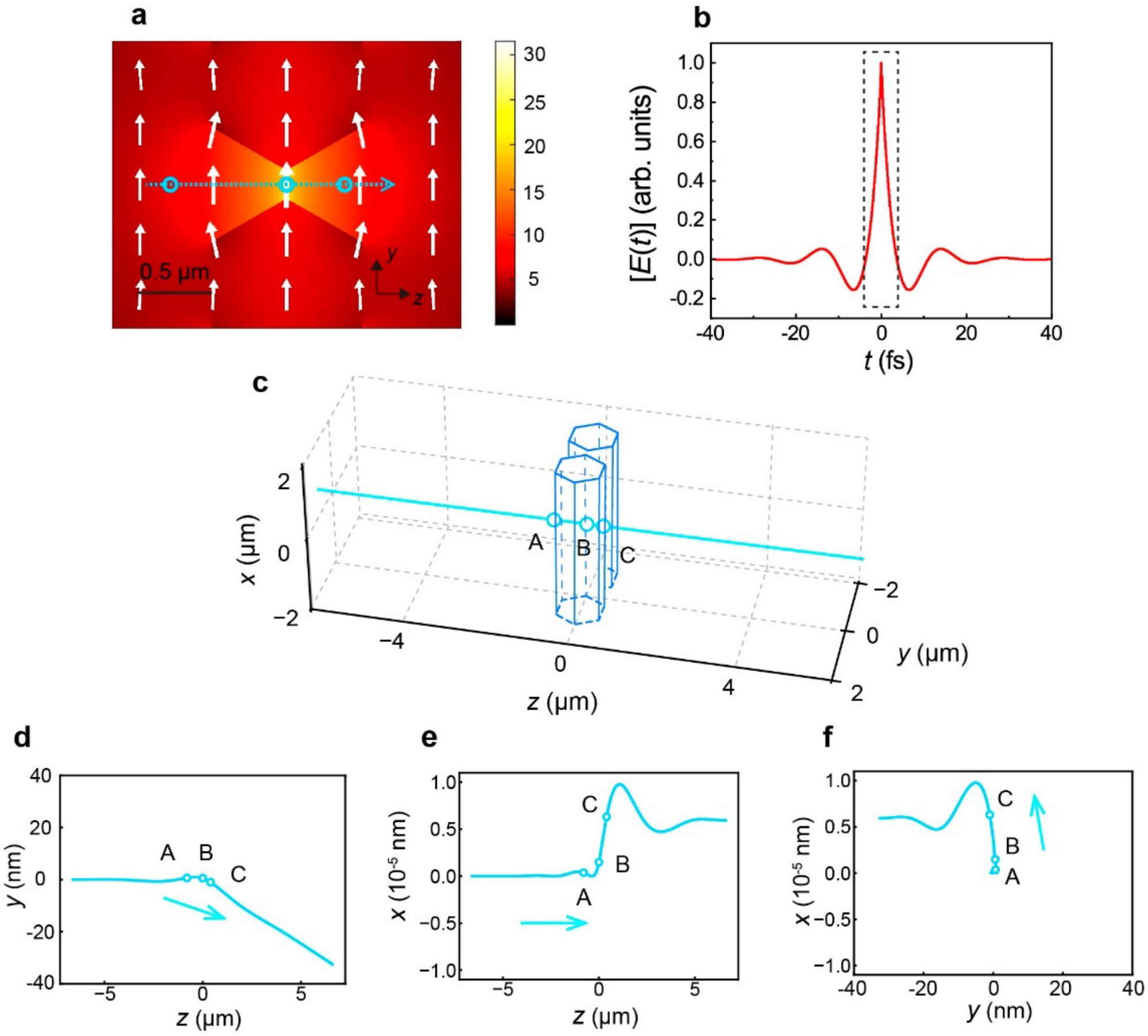

**Fig. S6| Calculation of the trajectory of a 2-MeV electron driven by a nano-slit optical field at 4.4-μm wavelength.** The nano-slit is constructed with two identical sapphire prisms with a diameter of 1.2 μm and a slit width of 200 nm. **a,** Calculated electric field of the optical driving field in the *y*-*z* plane. The orientation and size of the white arrow indicate the polarization and amplitude of the local field. The cyan dotted line arrow represents the direction of electron's trajectory, and the cyan circles represent electron's positions at *z*=-800 nm, 0 and 400 nm for A, B and C, respectively. **b,** *E*(*t*) of the nano-slit-mode optical driving field along *z* axis, noting that the spatial area in Fig. S6a corresponds to a narrow temporal region (black dashed rectangle). **c,** Schematic of the electron passing through the nano-slit. **d-f** Trajectories of the electron at cross-section planes of (**d**) *yz*, (**e**) *xz*, and (**f**) *xy*, respectively.

### 1.2.5 Compare the calculation between the ICS model and classical radiation

In the above calculations, the transverse electric field $\boldsymbol{E}(\boldsymbol{t},\boldsymbol{\theta})$ of an ICS pulse is obtained by Eq.17 from the well-known ICS model. On the other hand, from classical electrodynamics, the radiation field from an accelerated electron (here the electron is accelerated by the optical driving field) can be obtained as[4]

$$\boldsymbol{E}(\boldsymbol{t},\boldsymbol{\theta}) = \frac{e}{4\pi\varepsilon_0}\left\{\frac{\left[\hat{\boldsymbol{r}}\times[(\hat{\boldsymbol{r}}-\boldsymbol{\beta})\times\dot{\boldsymbol{\beta}}]\right]}{cr(1-\hat{\boldsymbol{r}}\cdot\boldsymbol{\beta})^3}\right\}, \quad (25)$$

where $\hat{\boldsymbol{r}}$ is the unit vector pointing from the electron to the observer (with *r* indicating the distance), and $\dot{\boldsymbol{\beta}}$ is the acceleration of electrons with

$$\dot{\boldsymbol{\beta}} = -\frac{e(\boldsymbol{E}_{\text{driving}}+\boldsymbol{V}_{\text{e}}\times\boldsymbol{B}_{\text{driving}})}{\gamma m_0} \tag{26}$$

To compare the two different models, using the same parameters in Fig. 3, we calculate the electric field of the scattered/radiated photons and its Fourier spectrum based on Eqs.18, 19, and 25, with the results given in Fig. S7.

It shows that, the results from the two models agree very well, with the same pulse width of 99 as and the peak frequency $\nu_{\text{peak}}$ of ~ 2.6 PHz (~ $T_{\text{peak}}$ of 0.38 fs).

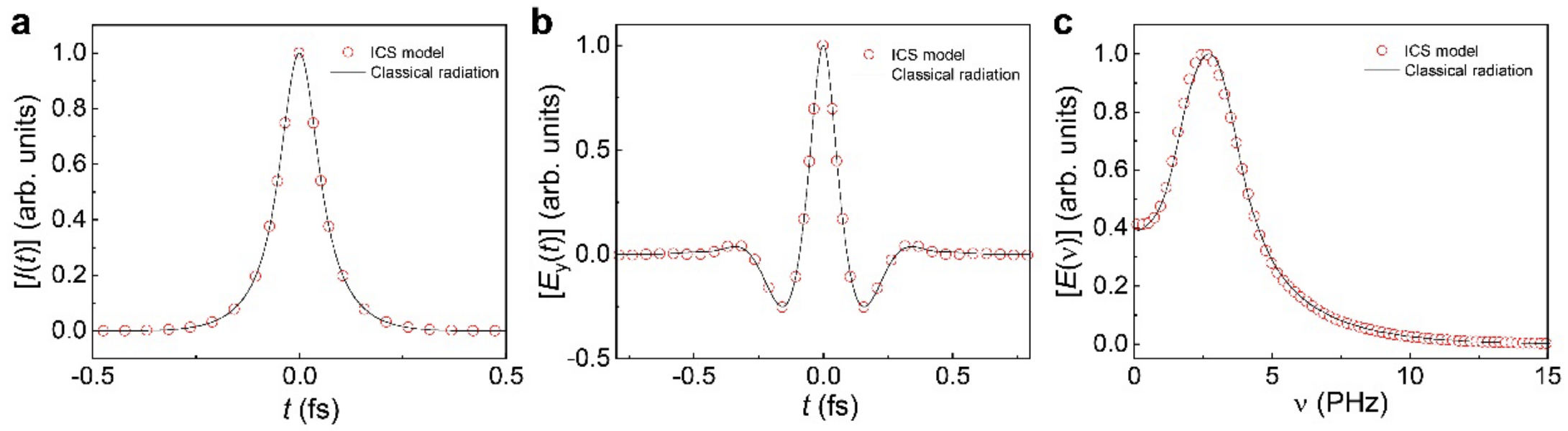

**Fig. S7| Comparison of electric fields calculated with ICS model (red circle) and classical radiation model (black solid line). a,** $I(t)$, **b,** $E_y(t)$ and **c,** $E(\nu)$.

## 2. Peak power density around optical damage threshold

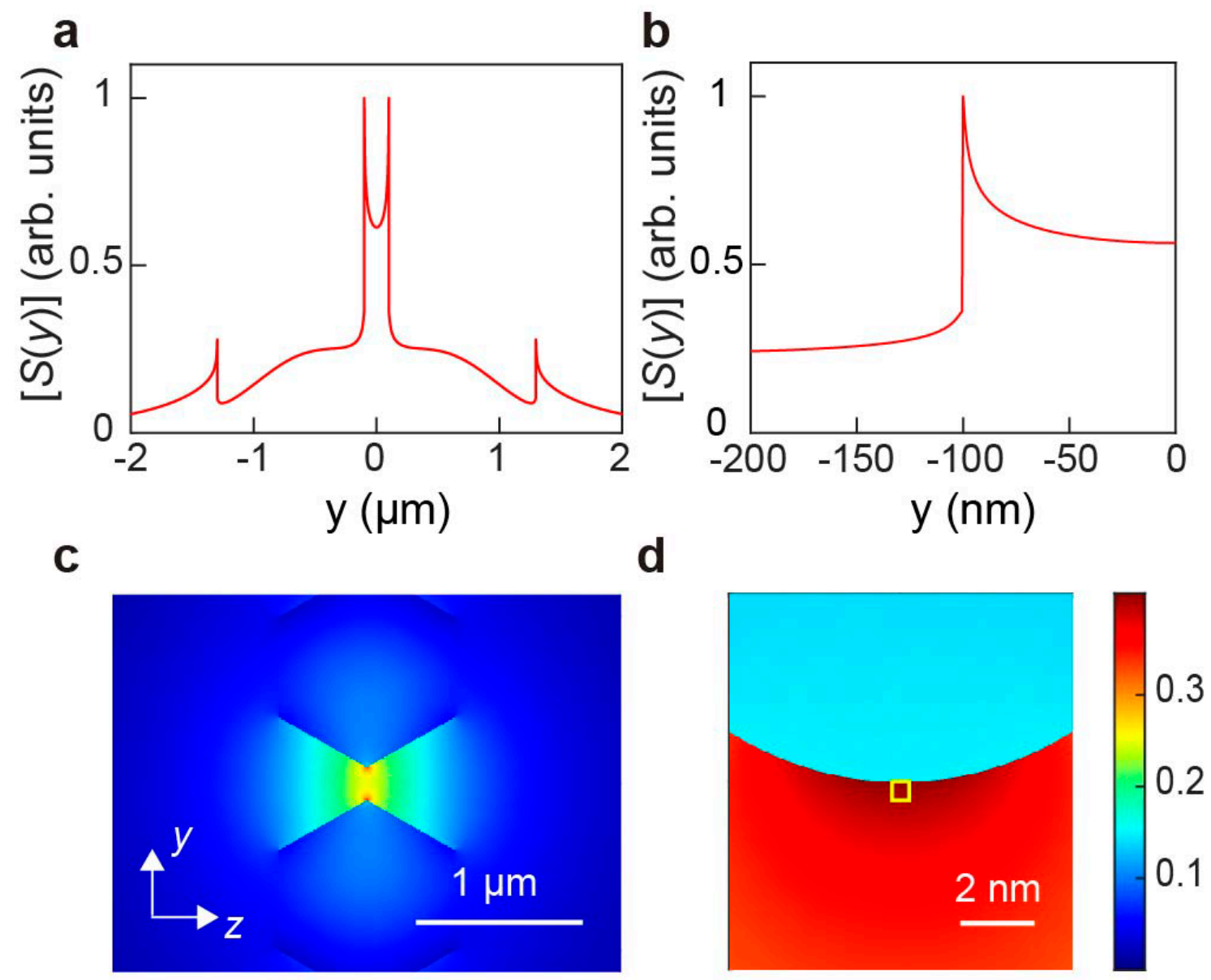

**Fig. S8| Derivation of the peak power density at the nano-slit edge of a coupled sapphire prism pair, with a prism diameter of 1.2 μm, a wavelength of 4.4 μm and a slit width of 200 nm. a,** Poynting vector along $y$ direction. **b,** Close-up view of the Poynting vector around the left edge of the nano-slit in (**a**). **c,** Poynting vector of the optical driving optical field in the $y$-$z$ plane. **d,** Close-up view of defining a 0.5 nm×0.5 nm rectangle at the vertex of the inner edge of the nano-slit.

The power densities at the center of the optical driving field and the inner surface of the nano-slit are obtained by the numerical calculation using COMSOL. Due to the abrupt refractive-index change at the edge of the nano-slit (i.e., edge of the sapphire prism), in the nano-slit mode, the maximum power density appears at both edges of the nano-slit (as shown in Fig. S8a-d). To calculate the power density around the edge, a small rectangle of 0.5 nm×0.5 nm is taken on the inner edge of the nano-slit (Fig. S8f), and the total power density in the small rectangle is approximately regarded as the optical power density at the edge of the nano-slit. When we define a ratio $\eta$ of the power in the small rectangle to the total power in the entire cross-section area, we obtain $\eta$~$1\times10^{-7}$. Then we use the same parameters in Fig. 3b in the main text, with a peak power of the driving laser of $4.4\times10^{6}$ W, we obtain a peak power in the small rectangle of ~ 0.44 W at the two vertically opposite vertices, corresponding to a power density of ~ $2\times10^{18}$ W/m$^2$ at the edge of the nano-slit.

## 3. Comparison of $\zeta$ between different methods

For easy comparison, calculated $\zeta$ of this and some previous works on attosecond or sub-cycle pulses are presented in Table 2 and Fig. S9.

Table 2 Experimental parameters and calculated $\zeta$ of some previous works on attosecond or sub-cycle pulse generation, based on high harmonic generation (HHG) or pulse coherent synthesis (CS) methods.

| Year of publication | Method | $\lambda_0$ (nm) | $\lambda_1$ (nm) | $\tau_1$ (as) | $\zeta$ | References |
|---|---|---|---|---|---|---|
| 2001 | HHG | 750 | 13.8 | 650 | 0.071 | [10] |
| 2006 | HHG | 750 | 34.5 | 130 | 0.885 | [11] |
| 2008 | HHG | 720 | 15.5 | 80 | 0.647 | [12] |
| 2011 | CS | 780 | 710 | 2100 | 1.13 | [13] |
| 2012 | CS | 800 | 760 | 3650 | 0.694 | [14] |
| 2012 | HHG | 750 | 9.55 | 67 | 0.475 | [15] |
| 2017 | HHG | 1800 | 3.76 | 53 | 0.237 | [16] |
| 2017 | HHG | 1800 | 6.9 | 43 | 0.535 | [17] |
| 2017 | CS | 2100 | 4200 | 12400 | 1.13 | [18] |
| 2020 | CS | 800 | 1400 | 2800 | 1.67 | [19] |
| 2024 | HHG | 1064 | 62 | 240 | 0.863 | [20] |

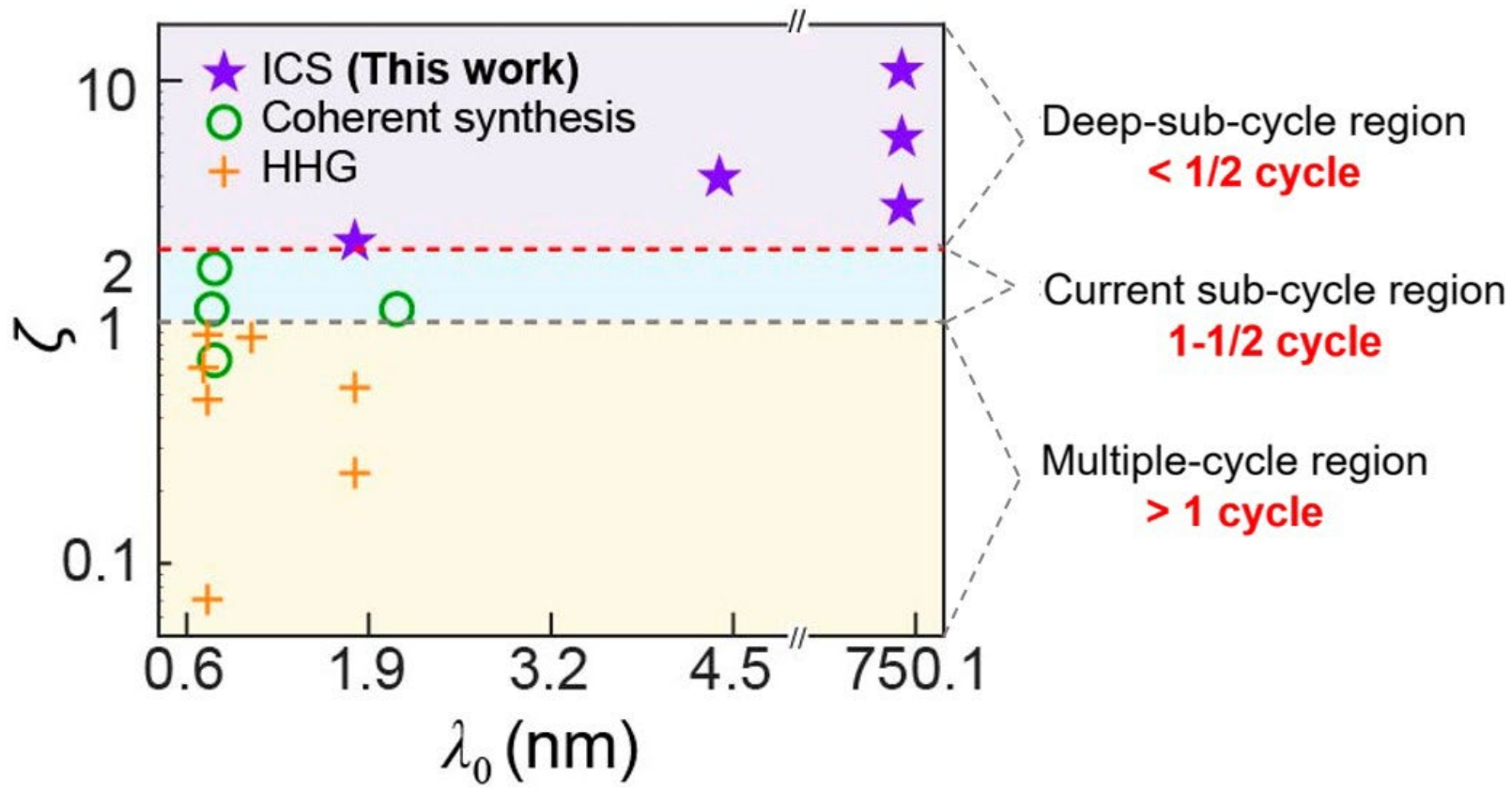

**Fig. S9| Calculated $\zeta$ of this and some previous works** (purple star: ICS; green circle: coherent synthesis; orange cross: HHG). ICS pulses are generated with same parameters in Fig. 2c, Fig. 3b, Fig. 4a and Figs. 4d-f.

## 4. Temporal evolution of the electrostatic field in an ICS pulse

The evolution and decay of the electrostatic field (i.e., the near field) in an ICS pulse can be quantitatively investigated by classical electrodynamic description of the radiating field of an electron

$$\boldsymbol{E}(\boldsymbol{t},\boldsymbol{\theta})=\frac{e}{4\pi\varepsilon_0}\left\{\frac{\left[\hat{\boldsymbol{r}}\times\left[(\hat{\boldsymbol{r}}-\boldsymbol{\beta})\times\dot{\boldsymbol{\beta}}\right]\right]}{cr(1-\widehat{\boldsymbol{r}\cdot}\boldsymbol{\beta})^3}+\frac{(1-\beta^2)(\hat{\boldsymbol{r}}-\boldsymbol{\beta})}{r^2(1-\hat{\boldsymbol{r}}\cdot\boldsymbol{\beta})^3}\right\} \tag{27}$$

where $r$ indicates the distance from the nano-slit center (i.e., $z$=0). The first term decays at a rate of $1/r$ and can radiate to the far field (corresponding to the purple solid line in Fig. 5); the second term decays at a rate of $1/r^2$, i.e., decays much more rapidly with increasing propagation distance (corresponding to the cyan dotted line in Fig. 5), which is also known as the near-field term. For reference, calculated $z$-dependent amplitude of the near-field part of $|E_y|$ is given in Fig. S10, clearly shown the quadratic decay behavior.

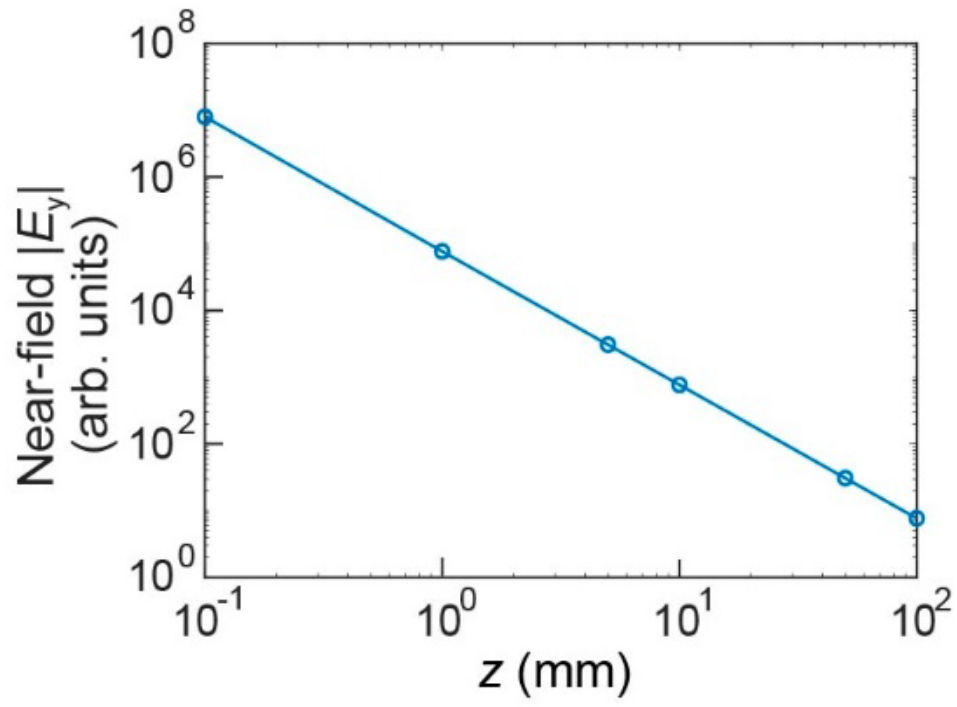

**Fig. S10| Calculated $z$-dependent amplitude of the near-field part of $|E_y|$,** with an optical driving field of 4.4-μm wavelength.